\documentclass[%
 reprint,
superscriptaddress,
bibnotes,
 amsmath,amssymb,
 aps,
prb,
]{revtex4-1}

\usepackage{graphicx}% Include figure files
\usepackage{dcolumn}% Align table columns on decimal point
\usepackage{bm}% bold math
\usepackage{xcolor}
\definecolor{OliveGreen}{rgb}{0,0.6,0}
\definecolor{auburn}{rgb}{0.43, 0.21, 0.1}
\definecolor{blue_violet}{rgb}{0.54, 0.17, 0.89}

\usepackage{appendix}
\usepackage{chngcntr}
\usepackage{etoolbox}
\usepackage{lipsum}

\AtBeginEnvironment{appendices}{%
\addcontentsline{toc}{section}{Appendices}
\counterwithin{figure}{section}
\counterwithin{equation}{section}
\counterwithin{table}{section}
}

\begin{document}

%\preprint{APS/123-QED}

\title{Steepest-Entropy-Ascent Quantum Thermodynamics Models in Materials Science}% Force line breaks with \\

\author{Ryo Yamada}
\email{ryo213@vt.edu}
\affiliation{Materials Science and Engineering Department, Virginia Polytechnic Institute and State University, Blacksburg, Virginia 24061, USA}
\author{Michael R. von Spakovsky}
\email{vonspako@vt.edu}
\affiliation{Center for Energy Systems Research, Mechanical Engineering Department, Virginia Polytechnic Institute and State University, Blacksburg, Virginia 24061, USA}
\author{William T. Reynolds, Jr.}
\email{reynolds@vt.edu}
\affiliation{Materials Science and Engineering Department, Virginia Polytechnic Institute and State University, Blacksburg, Virginia 24061, USA}

\date{\today}

\begin{abstract}
Steepest-entropy-ascent quantum thermodynamics, or SEAQT, is a unified approach of quantum mechanics and thermodynamics that avoids many of the inconsistencies that can arise between the two theories. Given a set of energy levels, i.e., energy eigenstructure, accessible to a given physical system, SEAQT predicts the unique kinetic path from any initial non-equilibrium state to stable equilibrium by solving a master equation that directs the system along the path of steepest entropy ascent. There are no intrinsic limitations on the length and time scales the method can treat so it is well-suited for calculations where the dynamics over multiple spacial scales need to be taken into account within a single framework. In this paper, the theoretical framework and its advantages are described, and several applications are presented to illustrate the use of the SEAQT equation of motion and the construction of a simplified, reduced-order, energy eigenstructure. 
\end{abstract}

\maketitle

\section{\label{chap2_sec:level1}Introduction}
Mechanics and equilibrium thermodynamics overlap extensively in computational materials science, but they have different origins. Quantum and classical mechanics describe non-entropic phenomena through a fundamental description of particle (and wave) behavior based upon Schr\"{o}dinger's or Newton's equation of motion. Thermodynamics is concerned with stable equilibria and provides a phenomenological description of matter derived from the first and second laws of thermodynamics. Because mechanics and thermodynamics developed independently and from different starting points, there are well-known conceptual incompatibilities between the two frameworks \cite{maddox1985uniting}. 

An intriguing theory that reconciles these incompatibilities appeared almost 40 years ago \cite{hatsopoulos1976-I,hatsopoulos1976-IIa,hatsopoulos1976-IIb,hatsopoulos1976-III,beretta2005generalPhD}. Its mathematical framework, which is now called steepest-entropy-ascent quantum thermodynamics (SEAQT), has developed extensively over the intervening years (e.g., see references \cite{beretta1984quantum, beretta1985quantum, beretta2006nonlinear, beretta2009nonlinear, beretta2014steepest, von2014some, montefusco2015essential, cano2015steepest, smith2016comparing, beretta2017steepest, li2016steepest, li2016generalized, li2016modeling, li2016steepest2, li2017study, li2018multiscale, li2018steepest, yamada2018method, yamada2018kineticpartI, yamada2018kineticpartII, yamada2018magnetization}). In the SEAQT theoretical framework, energy and entropy are used as fundamental state variables (as does classical thermodynamics), but entropy is interpreted as a measure of energy load sharing among available energy eigenlevels rather than as a statistical property of a statistical ensemble.  In addition, SEAQT postulates that the time-evolution of an isolated system maximizes the rate of entropy production at every instant of time. The particular path that satisfies this postulate is determined by a unique master equation called the SEAQT equation of motion, which directs the system along the path of steepest entropy ascent. 

The steps required to apply the SEAQT framework to materials-related problems are illustrated in this paper through several solid-state applications. By way of introduction, the SEAQT model is first compared and contrasted with common computational approaches in Section\;\ref{chap2_sec:level2}. In Sec.\;\ref{chap2_sec:level3}, the SEAQT equation of motion is derived for the case of an isolated system and for interacting systems. In Sec.\;\ref{chap2_sec:level4}, the issues associated with constructing an energy eigenstructure (a set of energy levels) are described for solids, and then a method for building a simplified energy eigenstructure (a so-called ``pseudo-eigenstructure") is presented to address these issues. In Sec.\;\ref{chap2_sec:level5}, the SEAQT model is demonstrated using a simple model system and then a ferromagnetic spin system with a focus on the use of the
SEAQT equation of motion and the construction of the pseudo-eigenstructure. Finally, the salient features and advantages of the SEAQT model are noted in Sec.\;\ref{chap2_sec:level6} along with some future directions for study.

\section{\label{chap2_sec:level2}Advantages of the SEAQT Model }

\subsection{\label{chap2_sec:level2_1}Mechanics and Thermodynamics}
\begin{figure}
\begin{center}
\includegraphics[scale=0.5]{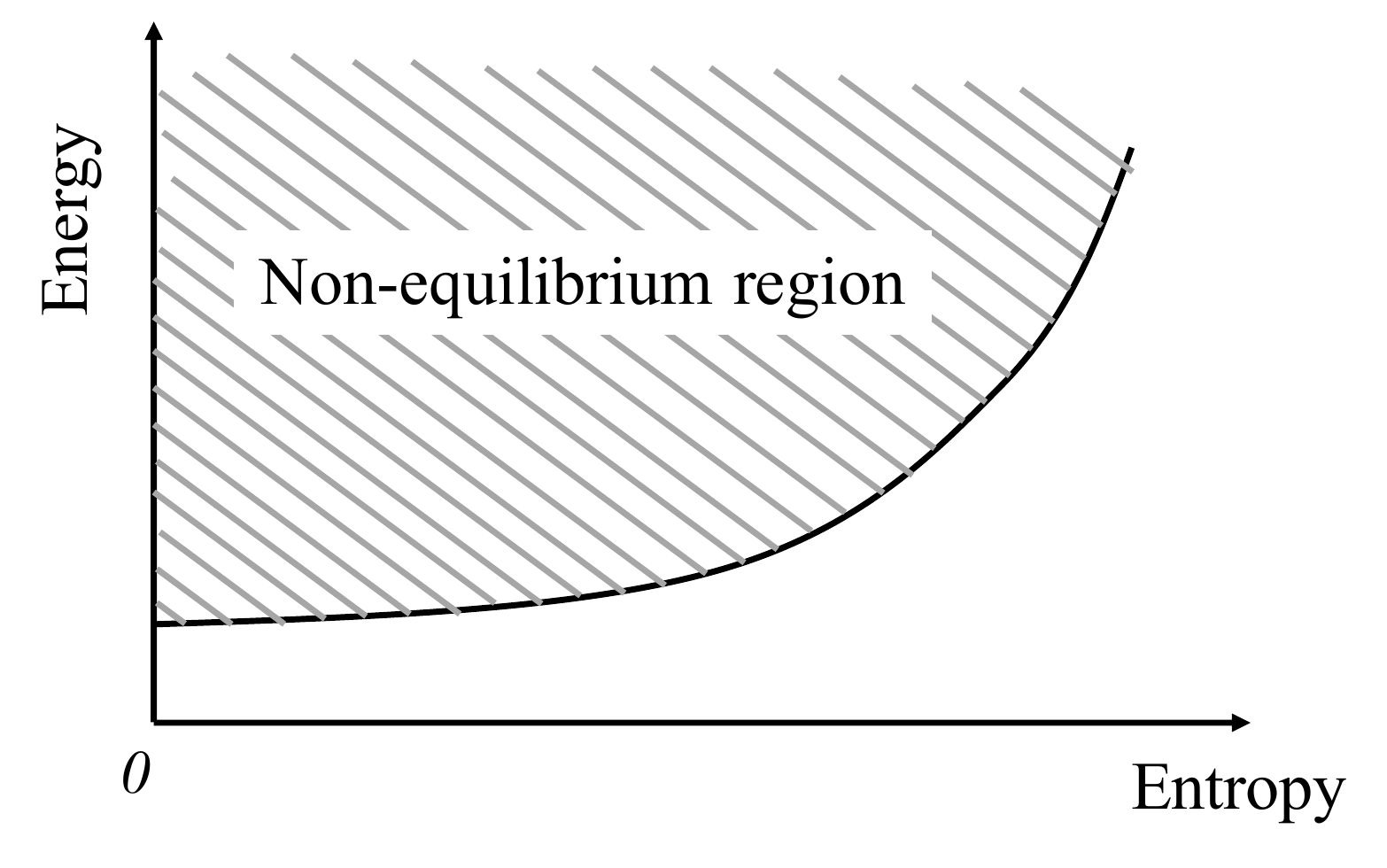}
\caption{\label{fig2:E_S_diagram} A schematic energy--entropy (\textit{E}--\textit{S}) diagram for a system with constant volume, $V$, and number of particles, $N$. The bounding curve represents the stable equilibrium states in the in the \textit{E}--\textit{S} plane described by equilibrium thermodynamics, and the vertical axis is a non-entropic line that represents the domain of mechanics in this plane. The cross-hatched area is the non-equilibrium region that is not strictly described either by mechanics or by thermodynamics.   }
\end{center}
\end{figure}

The energy--entropy (\textit{E}--\textit{S}) diagram (Fig.\;\ref{fig2:E_S_diagram}), which is a two-dimensional cut in the \textit{E}--\textit{S} plane of the hypersurface of all stable equilibrium states for a given system, helps clarify where the mechanics and equilibrium thermodynamic approaches are valid. While mechanics describes non-entropic states corresponding to the vertical axis of Fig.\;\ref{fig2:E_S_diagram}, classical thermodynamics is largely limited to the stable equilibria represented by the bounding curve in the figure.

A variety of material properties can be calculated reliably in the non-entropic region using first principle methods for solving Schr\"{o}dinger-like equations (e.g., the Kohn-Sham equations of density functional theory), but these methods cannot be employed directly at the finite temperatures of the entropic region. In order to determine properties at finite temperatures, quantum statistical mechanics is often combined with density functional theory where the most probable state is explored by searching the minimum free energy. Quantum statistical mechanics has had much success describing solid-state phenomena such as magnetic transitions, gas--liquid transitions, and order--disorder transformations \cite{kittel1980thermal,girifalco2003statistical}. However, it introduces unphysical assumptions by assuming a heterogeneous ensemble (Appendix\;\ref{chap2_sec:level7_1}) and its applicability is limited to the stable equilibrium region and does, thus, not apply to the non-equilibrium region (the cross-hatched area in Fig.\;\ref{fig2:E_S_diagram}).

There are a number of ways to combine quantum mechanics with thermodynamics to describe non-equilibrium time-evolution processes at the quantum scale \cite{smith2012intrinsic}. For example, using a nonlinear time-dependent Schr\"{o}dinger equation of motion \cite{doebner1992general,schuch2010pythagorean} with an added frictional term or Markovian and non-Markovian quantum master equations \cite{gemmer2004quantum,gemmer2009quantum,zurek1994decoherence} where so-called ``dissipative open systems'' are assumed are two such ways that this can be done. Unfortunately, as recently pointed out, these approaches are plagued by inconsistencies in descriptions such as the definition of state, which is  different in each of the approaches \cite{smith2012intrinsic}. It is simply noted here without dwelling on these inconsistencies that the SEAQT framework provides an alternative approach for unifying quantum mechanics and thermodynamics that does not introduce any intrinsic inconsistencies. Additional details can be found in reference \cite{smith2012intrinsic}.

\subsection{\label{chap2_sec:level2_2}Multiscale calculations in materials science}
Computational investigations of materials cover a broad range of length and time scales. Macroscopic material properties generally depend to some extent on the underlying atomistic, microscopic, and mesoscopic behavior. For example, the deformation behavior of a structural steel component depends not only upon the geometry of the component but also on the steel microstructure and its dependence upon the local plastic deformation zones, which, in turn, depend upon the atomic bonding of the constituent atoms. 

\begin{figure}
\begin{center}
\includegraphics[scale=0.52]{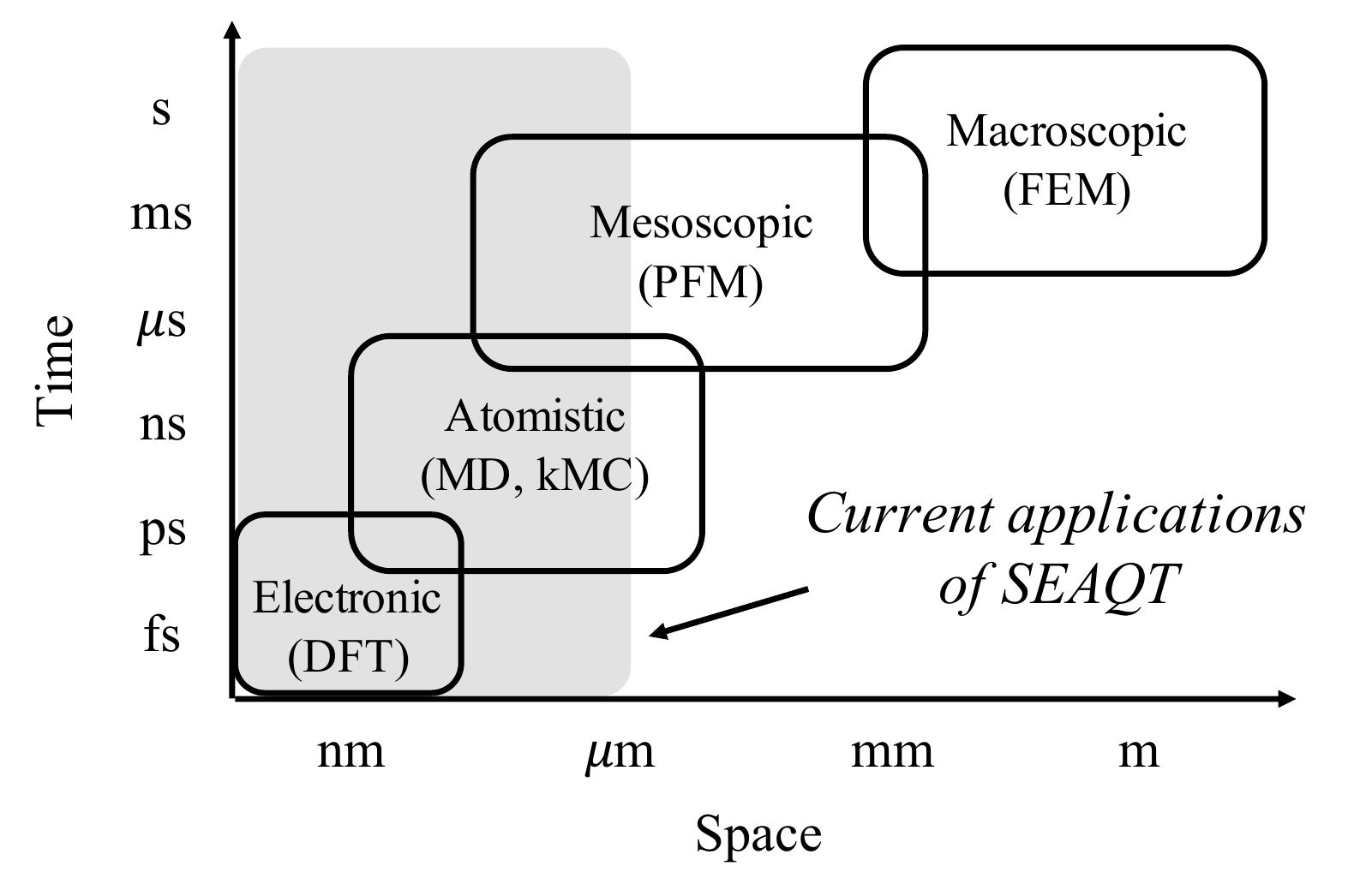}
\caption{\label{fig2:multi_scale_calculation} Common computational methods and the different time and length scales for their application in materials science \cite{onodera2014recent}. The acronyms shown are Finite Element Method (FEM), Phase Field Model (PFM), Molecular Dynamics (MD), kinetic Monte Carlo (kMC), and Density Functional Theory (DFT). The range of scales over which SEAQT has been applied to date is indicated by the gray region; there are no intrinsic limitations that prevent it from being extended over larger spatial scales.}
\end{center}
\end{figure}

Approaches suitable for calculating material properties apply to different length and time scales (Fig.\;\ref{fig2:multi_scale_calculation}). For instance, in the above example of deformation behavior, macroscopic strains are calculated using the finite element method \cite{bathe2007finite,dhatt2012finite}, microstructure evolution at the mesoscopic spatial scale with phase field models \cite{chen2002phase,moelans2008introduction}, atomic displacements with molecular dynamics simulations \cite{binder2004molecular} and kinetic Monte Carlo simulations, and bonding-level behavior with electronic structure calculations \cite{voter2007introduction,lesar2013introduction}. 

Each computational method is quite successful when applied over the length and time scales for which it was developed, but extending them to other length/time scales is problematic. To overcome these difficulties, computational methods have been combined synergistically such that the time-dependence of a property is calculated in a larger-scale computational model with parameters/data determined from smaller-scale methods in a ``constitutive approach" \cite{weinan2011principles}. For example, a deformation process can be simulated by calculating atomistic parameters with molecular dynamics \cite{hoyt2002atomistic} or Monte Carlo simulations \cite{vaithyanathan2004multiscale} and then passing them to a phase field model that calculates the microstructure  \cite{yamanaka2008coupled,fromm2012linking}, which is subsequently passed to a finite element method that simulates the deformation process. 
%They may be reasonable if the smaller sale dynamics equilibrate much faster than the larger scale and/or the parameters are invariable in the larger scale calculation. However, this is not always true. For instance, since the relaxation time of phonon and spinodal decomposition would not be so different  and the microstructure would be altered during the deformation process. , so the atomic diffusivity calculated in MD should not be equilibrated in the spinodal decomposition simulated in PFM).   
Although the constitutive approach connects different length scales, the dynamics at smaller scales are usually ignored by the larger scales. As pointed out in reference \cite{weinan2011principles}, although the constitutive approach may be adequate for a simple system, its applicability to a complex system is questionable, because complex interactions among scales are possible and many parameters would be required to represent them. Furthermore, parameters/data in the constitutive relation are calculated ignoring the effect of larger-scale phenomena by assuming a homogeneous system \cite{weinan2011principles}. Therefore, in order to reliably describe behavior over multiple scales, it is desirable to combine methods that take into account the dynamics at each scale and mutually update data during the entire time-evolution process. This is difficult with existing methods because the state variables and governing equations differ from one scale to the next and converting variables and using different governing equations becomes very problematic \cite{li2018multiscale}.    

The SEAQT framework has the potential to improve this situation. Unlike the computational methods described above, the SEAQT framework uses energy and entropy as its fundamental state variables and the time-evolution of a system is determined from the SEAQT equation of motion based on the principle of steepest entropy ascent at each instant of time. Since energy and entropy can be defined for any state in any system regardless of scale and the equation of motion is based upon quantum mechanics without resort to the near/local equilibrium assumptions, the framework applies to any state at all length and time scales. Thus, it is able to describe physical phenomena and their couplings at all length and time scales within a single theoretical framework \cite{li2018multiscale}.  

The SEAQT framework has several additional distinguishing characteristics relative to conventional computational models. They are as follow:
\begin{itemize}
\item MD is limited to high temperatures (above the Debye temperature) because it is based on classical mechanics, while SEAQT is equally valid at all temperatures. In addition, MD models require an artificial term in the Hamiltonian when a system interacts with a heat reservoir \cite{lesar2013introduction}, while there is no need to introduce arbitrary terms in the Hamiltonian with the SEAQT approach since the framework is based on a fundamental and not a phenomenological description.

\item Whereas the PFM is most appropriate for near-stable equilibrium states because the time-evolution process is determined by a master equation (e.g., the Cahn-Hilliard equation and the Allen-Cahn equation \cite{balluffi2005kinetics}) that is derived assuming small deviations from equilibrium, the SEAQT framework requires no such restriction, because the SEAQT equation of motion does not require the near/local equilibrium assumption.

\item While the kMC method needs to identify all possible discrete events that can take place at each instant of time, the kinetic path in SEAQT is determined by merely solving the SEAQT equation of motion (a set of first-order, ordinary differential equations). Thus, the computational burden associated with the SEAQT framework is small compared to that for kMC (as well as the other methods described here). Moreover, the stochastic framework in kMC can make it difficult to extract physical insights from the simulations without a statistical analysis of multiple computational experiments.
\end{itemize}

\section{\label{chap2_sec:level3}SEAQT equation of motion}
\begin{figure}
\begin{center}
\includegraphics[scale=0.42]{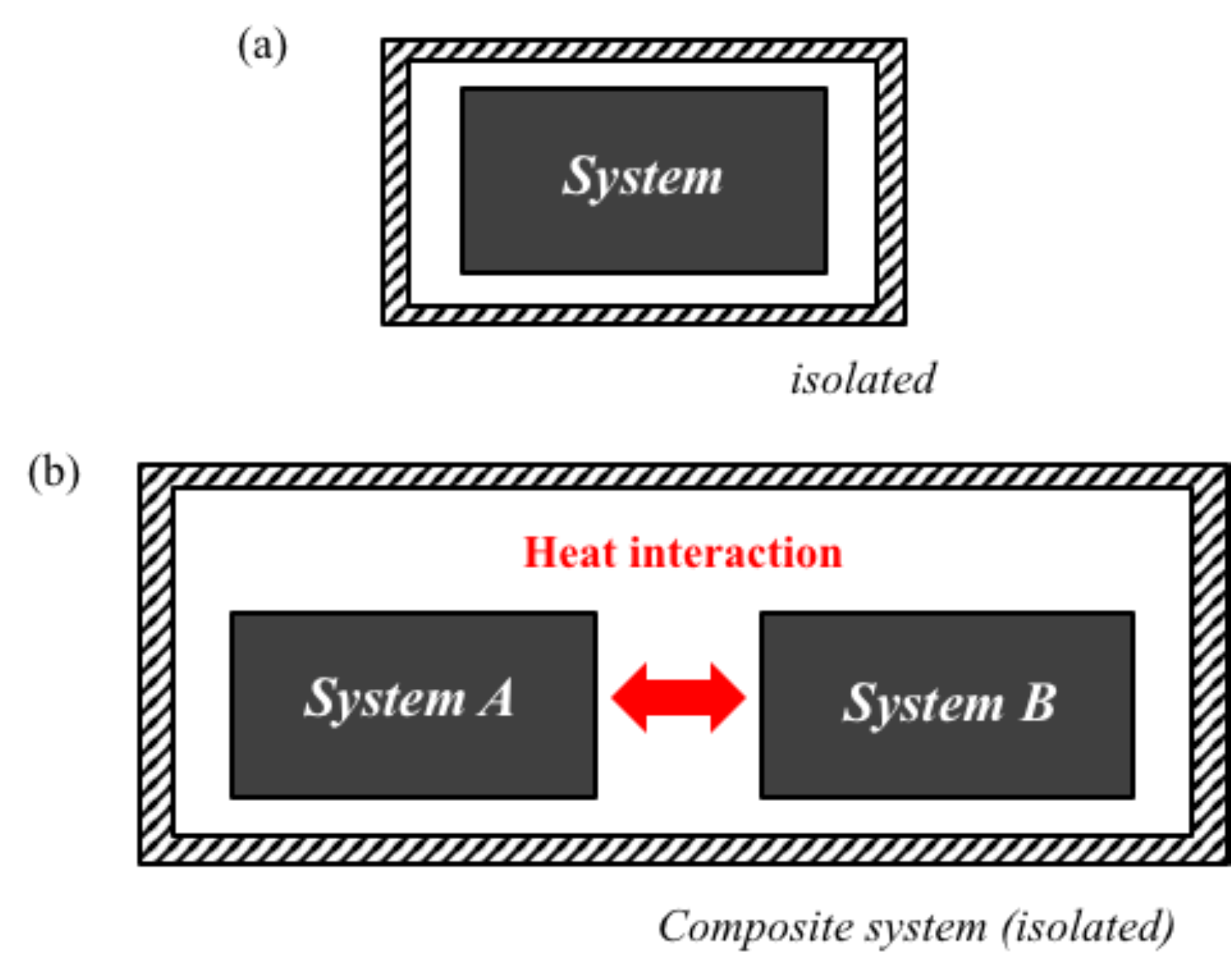}
\caption{\label{fig2:isolated_systems} Schematic descriptions of the isolated systems under consideration: (a) the simple isolated system considered in Sec.\;\ref{chap2_sec:level3_1}, and (b) the isolated system with two subsystems that exchange energy in a heat interaction in Sec.\;\ref{chap2_sec:level3_2}.   }
\end{center}
\end{figure}

The SEAQT equation of motion is based on the steepest-entropy-ascent principle using energy and entropy as the basic state variables, and it has been demonstrated that the equation of motion recovers the Boltzmann transport equations in the near-equilibrium limit \cite{li2018steepest}. Here, the SEAQT equation of motion is derived for an isolated system and for an isolated composite system that contains two interacting systems that exchange energy in a heat interaction (Fig.\;\ref{fig2:isolated_systems}).

\subsection{\label{chap2_sec:level3_1}Isolated system}
A typical quantum mechanics equation of motion, such as a Schr\"{o}dinger-like equation, only describes a subset of reversible processes (i.e., those involving non-entropic phenomena). The SEAQT equation of motion, on the other hand, adds a postulated dissipative term to the time-dependent Schr\"{o}dinger equation that makes it possible to describe both reversible and irreversible processes. This equation for a simple (as opposed to general) quantum system is written as \cite{beretta1984quantum,beretta1985quantum,beretta2006nonlinear,beretta2009nonlinear} 
\begin{equation}
\frac{d\hat{\rho}}{dt}=\frac{1}{i \hbar}[\hat{\rho},\hat{H}] + \frac{1}{\tau(\hat{\rho})}\hat{D}(\hat{\rho}) \; , \label{eq2:equation_of_motion}
\end{equation}
where $\hat{\rho}$ is the density operator, $t$ the time, $\hbar$ the reduced Planck constant, $\hat{H}$ the Hamiltonian operator, $\tau$ the relaxation time, and $\hat{D}$ the dissipation operator. The left-hand side of the equation and the first term on the right corresponds to the time-dependent von Neumann equation (or Schr\"{o}dinger equation), and the second term on the right is the dissipation term --- an irreversible contribution that accounts for relaxation processes in the system. The density operator, $\hat{\rho}$, includes all the information about the state of the system. Its use allows SEAQT to unify quantum mechanics and thermodynamics into a consistent theoretical framework \cite{hatsopoulos1976-I,hatsopoulos1976-IIa,hatsopoulos1976-IIb,hatsopoulos1976-III}. 

When there are no quantum correlations between particles, $\hat{\rho}$ is diagonal in the Hamiltonian eigenvector basis \cite{li2016generalized,li2016modeling,li2017study} and $\hat{\rho}$ and $\hat{H}$ commute, i.e., $[\hat{\rho},\hat{H}]=0$. Under this circumstance, the SEAQT equation of motion, Eq.\;(\ref{eq2:equation_of_motion}), reduces to \cite{beretta2006nonlinear,beretta2009nonlinear,li2016steepest}
\begin{equation}
\frac{dp_j}{dt}= \frac{1}{\tau(\bm{p})} D_j(\bm{p}) \; , \label{eq2:equation_of_motion_second_term}
\end{equation}
where the $p_j$ are the diagonal terms of $\hat{\rho}$, each of which represents the occupation probability in the $j^{th}$ energy eigenlevel, ${\epsilon}_j$, and $\bm{p}$ denotes the vector of all the $p_j$. (Since the contribution of quantum correlations would be quite small for most material properties, the form of the SEAQT equation of motion shown in Eq.\;(\ref{eq2:equation_of_motion_second_term}) is employed hereafter.) The dissipation term, $D_j(\bm{p})$, can be derived via either a variational principle \cite{beretta2006nonlinear} or via the use of a manifold \cite{beretta2006nonlinear,beretta2009nonlinear,li2016steepest} with the postulate that the time-evolution of a system follows the direction of steepest entropy ascent constrained by appropriate conservation laws. Here, the derivation of the SEAQT equation of motion is briefly described using the mathematical technique of a manifold constrained by the conservation of energy and conservation of the occupation probabilities. 

\begin{figure}
\begin{center}
\includegraphics[scale=0.42]{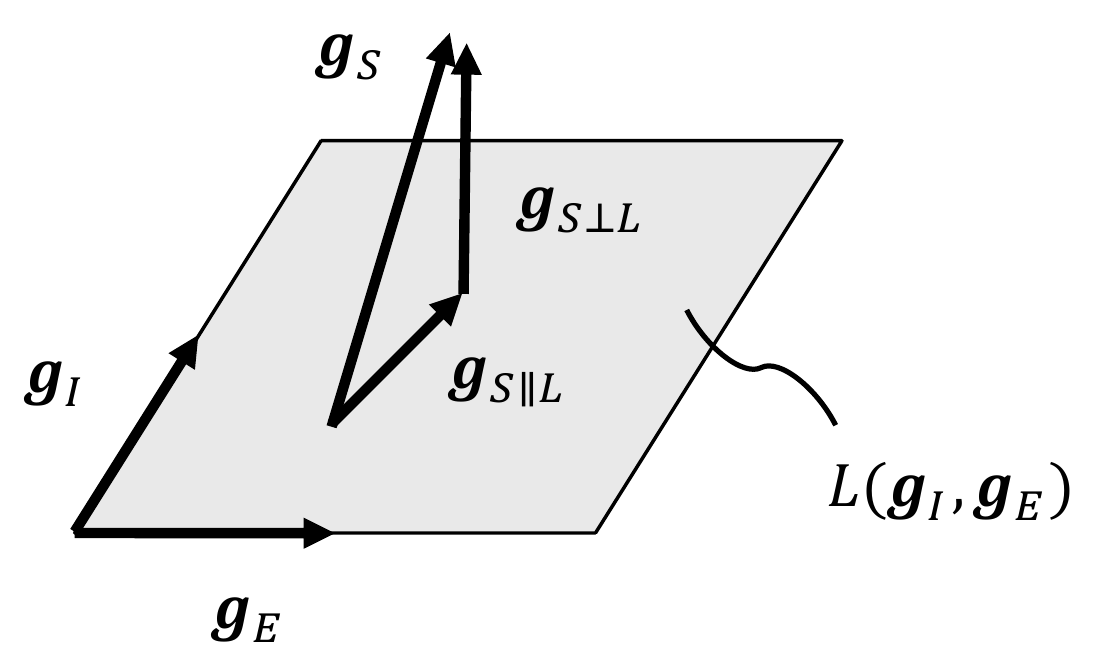}
\caption{\label{fig2:manifold} Geometric representation of the steepest-entropy-ascent direction constrained by the conservation of occupation probabilities and the energy \cite{beretta2005generalPhD,smith2012intrinsic}. The gradients $\bm{g}_I$, $\bm{g}_E$, and $\bm{g}_S$ are, respectively, the gradients of the occupation probabilities, energy, and entropy in state space, and $L(\bm{g}_I, \bm{g}_E)$ is the manifold spanned by $\bm{g}_I$ and $\bm{g}_E$. While $\dot{\bm{x}}$ would be in the direction of $\bm{g}_S$ for an unconstrained process, it must be orthogonal to the manifold for maximum entropy generation, i.e., $\bm{g}_{S \bot L(\bm{g}_I, \bm{g}_E)}$, in order to conserve the occupation probabilities and the energy \cite{beretta2006nonlinear}. }
\end{center}
\end{figure}

For the purpose of deriving the dissipation term, $D_j(\bm{p})$, the square root of the probability distribution, $x_j=\sqrt{p_j}$, is employed (as is done in references \cite{beretta2006nonlinear,beretta2009nonlinear,li2016steepest}). Using $x_j$, the summation of the occupation probabilities and the expected energy and entropy of a system are written as \cite{li2016steepest}
\begin{equation}
\begin{split}
 & \quad \quad \quad \quad   I = \sum_i p_i = \sum_i x_i^2 \;  \\
 & \quad \quad  E =  \langle e \rangle = \sum\limits_{i} \epsilon_i p_i = \sum\limits_{i} \epsilon_i x_i^2  \;  \\
 S = & \langle s \rangle =  - \sum\limits_{i} p_i \mathrm{ln} \left( \frac{p_i}{g_i} \right) = - \sum\limits_{i} x_i^2 \mathrm{ln} \left( \frac{x_i^2}{g_i} \right)  \; 
 , \label{eq2:extensive_property_isolated}
\end{split}
\end{equation}
where $g_j$ is the degeneracy of the energy eigenlevel $\epsilon_j$. The von Neumann formula for entropy is used in the last line of Eq.\;(\ref{eq2:extensive_property_isolated}) because it satisfies all the characteristics required by thermodynamics \cite{gyftopoulos1997entropy,cubukcu1993thermodynamics} (the quantum Boltzmann entropy formula is discussed in Appendix\;\ref{chap2_sec:level7_1}). The gradients of each property in state space are then expressed as
\begin{equation}
\begin{split}
& \quad \quad \quad  \; \bm{g}_I = \sum_i \frac{\partial I}{\partial x_i} \hat{e}_i =  \sum_i 2x_i \hat{e}_i  \;  \\
& \quad \quad \quad  \bm{g}_E  = \sum_i \frac{\partial E}{\partial x_i} \hat{e}_i =  \sum_i 2 \epsilon_i x_i \hat{e}_i  \;  \\
\bm{g}_S & = \sum_i \frac{\partial S}{\partial x_i} \hat{e}_i = - \sum_i 2x_i \left[ 1 + \mathrm{ln} \left( \frac{x_i^2}{g_i} \right) \right] \hat{e}_i  \; 
 , \label{eq2:gradient_extensive_property_isolated}
\end{split}
\end{equation}
where $\hat{e}_i$ is the unit vector for component, $i$, i.e., the $i^{th}$ eigenlevel. Since $I=1$ and $E=$\;constant, the time-evolution of state, $\dot{\bm{x}}$\;(=$d\bm{x}/dt$), must be orthogonal to the manifold spanned by $\bm{g}_I$ and $\bm{g}_E$. That is, $\bm{\dot{g}}_I$\;(=$d\bm{g}_I/dt$) and $\bm{\dot{g}}_E$\;(=$d\bm{g}_E/dt$) must be zero (see Fig.\;\ref{fig2:manifold}). Therefore, the time-evolution is given by the solution of \cite{beretta2006nonlinear,beretta2009nonlinear,li2016steepest}
\begin{equation}
\begin{split}
 \frac{d\bm{x}}{dt} & = \frac{1}{\tau(\bm{x})} \bm{g}_{S \bot L(\bm{g}_I, \bm{g}_E)}  \\
& = \frac{1}{\tau(\bm{x})} \frac{\begin{vmatrix} 
\bm{g}_S & \bm{g}_I & \bm{g}_E \\
(\bm{g}_S, \bm{g}_I) & (\bm{g}_I, \bm{g}_I) & (\bm{g}_E, \bm{g}_I) \\
(\bm{g}_S, \bm{g}_E) & (\bm{g}_I, \bm{g}_E) & (\bm{g}_E, \bm{g}_E)
\end{vmatrix}}{\begin{vmatrix} 
(\bm{g}_I, \bm{g}_I) & (\bm{g}_E, \bm{g}_I) \\
(\bm{g}_I, \bm{g}_E) & (\bm{g}_E, \bm{g}_E) 
\end{vmatrix}} 
\; , \label{eq2:equation_of_motion_second_term2}
\end{split}
\end{equation}
where $L(\bm{g}_I, \bm{g}_E)$ is the manifold spanned by $\bm{g}_I$ and $\bm{g}_E$ and $\bm{g}_{S \bot L(\bm{g}_I, \bm{g}_E)}$ is the perpendicular component of the gradient of the entropy, $\bm{g}_S$, to the manifold, which is written in an explicit form using the theory of Gram determinants \cite{beretta2006nonlinear} (the notation $(\cdot \, ,\cdot)$ represents the scalar product of two vectors). The explicit form of the SEAQT equation of motion for this case is then written as \cite{beretta2006nonlinear,beretta2009nonlinear,li2016steepest}
\begin{equation}
\frac{dp_j}{dt^*}=\frac{\begin{vmatrix} 
-p_j \mathrm{ln} \frac{ p_j }{g_j} & p_j & \epsilon_jp_j \\
\langle s \rangle & 1 & \langle e \rangle \\
\langle es \rangle & \langle e \rangle & \langle e^2 \rangle
\end{vmatrix}}{\begin{vmatrix} 
1 & \langle e \rangle \\
\langle e \rangle & \langle e^2 \rangle 
\end{vmatrix}} \; ,  \label{eq2:equation_of_motion_simplified}
\end{equation}
where 
\[
\begin{array}{c c}
\langle e^2 \rangle = \sum\limits_{i} \epsilon_i^2 p_i \; , \;\;\;
&
\langle es \rangle = - \sum\limits_{i} \epsilon_i p_i \mathrm{ln} \frac{ p_i }{g_i}  \; ,
\end{array}
\]
and $t^*$\;($\;=\frac{t}{\tau (\bm{p})}$) is the dimensionless time and $\tau (\bm{p})$ a relaxation time. In Eq. (\ref{eq2:equation_of_motion_simplified}), the time-dependent trajectory of state evolution, $p_j(t^*)$ is expressed in terms of a dimensionless time rather than in terms of the real time, $t$. The two kinds of time are distinguished by using the term `kinetics' to refer to processes expressed in terms of $t^*$ and `dynamics' to refer to processes expressed in terms of $t$. Thus, the `kinetics' establishes the unique thermodynamic path along which the state of the system evolves in state space (e.g., Hilbert space), while $\tau$ determines the speed at which the system evolves along this path, i.e., the so-called `dynamics'. A detailed discussion of this distinction can be found in references \cite{li2016steepest,li2016generalized}.

The derivation of the SEAQT equation of motion can be extended to include additional conservation conditions, e.g., the number of particles \cite{li2016steepest2}, the volume \cite{li2016modeling}, and the magnetization \cite{yamada2018magnetization}. The SEAQT equation of motion with constant magnetization is shown in Sec.\;\ref{chap2_sec:level5_2}.

\subsection{\label{chap2_sec:level3_2}Heat interaction between systems}
The SEAQT equation of motion was formally derived in the previous section (Sec.\;\ref{chap2_sec:level3_1}) for an isolated system  but can also be extended to interacting systems by treating them as interacting systems within a larger, isolated composite system \cite{li2016generalized,li2016steepest2} (see Fig.\;\ref{fig2:isolated_systems}\;(b)). (Hereafter, we call the interacting systems ``subsystems" within the composite.) Furthermore, if one of the subsystems is much larger than the other, the bigger subsystem can be treated as a reservoir and the SEAQT equation of motion for a system interacting with a reservoir can be formulated as well \cite{li2016generalized,li2016steepest2}. 

To derive the SEAQT equation of motion for two (sub)\;systems, $A$ and $B$, interacting via a heat interaction, three quantities in the composite system must be conserved: the energy of the overall composite system and the occupation probabilities in each subsystem. In this case, the manifold can be expressed as $L=L(\bm{g}^A_{I}, \bm{g}^B_{I}, \bm{g}_E)$. The equation of motion for each subsystem takes the form \cite{li2016generalized} 
\begin{equation}
\frac{dp^A_j}{dt^*}=\frac{\begin{vmatrix} 
-p^{A}_j \mathrm{ln} \frac{p^A_j}{g^A_j} & p^A_j & 0 & \epsilon^A_j p^A_j   \\
\left< s \right>^A & 1 & 0 & \left< e \right>^A  \\
\left< s \right>^B & 0 & 1 & \left< e \right>^B  \\
\left< es \right> & \left< e \right> ^A & \left< e \right> ^B & \left< e^2 \right> \\
\end{vmatrix}}{\begin{vmatrix} 
1 & 0 & \left< e \right>^A  \\
0 & 1 & \left< e \right>^B  \\
 \left< e \right>^A & \left< e \right>^B & \left< e^2 \right> \\
\end{vmatrix}} \; ,  \label{eq2:equation_of_motion_interacting}
\end{equation}
where $\left< \cdot \right>^{A\;(\mbox{\scriptsize or} B)}$ is the expectation value of a property in subsystem\;$A$ (or $B$), and $\left< \cdot \right>= \left< \cdot \right>^A + \left< \cdot \right>^B$ is the property in the composite system (only the equation of motion for system\;$A$ is shown above). Representing the cofactors of the first line of the determinant in the numerator by $C_1$, $C^A_2$, and $C_3$, Eq.\;(\ref{eq2:equation_of_motion_interacting}) can be expressed as \cite{li2016generalized} 
\begin{equation}
\begin{split}
 \; \frac{dp^A_j}{dt^*} &= p^A_j \left( - \mathrm{ln} \frac{p^A_j}{g^A_j} - \frac{C_2^A}{C_1} - \epsilon^A_j  \frac{C_3}{C_1}  \right)  \\
& = p^A_j \left[ (s^A_j -  \left< s \right>^A)  - (\epsilon^A_j - \left< e \right>^A) \frac{C_3}{C_1}   \right] \\
& = p^A_j \left[ (s^A_j -  \left< s \right>^A)  - (\epsilon^A_j - \left< e \right>^A) \beta  \right]  \; . \label{eq2:equation_of_motion_interacting_simplified}
\end{split}
\end{equation}
The factor $\beta$ is defined as $\beta \equiv C_3/C_1$ because it can be related to a temperature, $T$, as $\beta=\frac{1}{k_B T}$ using the concept of hypo-equilibrium states described in Appendix\;\ref{chap2_sec:level7_2}. Here, $k_B$ is Boltzmann's constant. In addition, $\beta$ is related to the mole fractions of the subsystems \cite{li2016generalized}. Therefore, when system\;$B$ of Fig.\;\ref{fig2:isolated_systems}\;(b) is much larger than system\;$A$ and viewed as a heat reservoir, Eq.\;(\ref{eq2:equation_of_motion_interacting_simplified}) is transformed into \cite{li2016generalized}
\begin{equation}
\frac{dp_j}{dt^*} = p_j \left[ (s_j -  \left< s \right>)  - (\epsilon_j - \left< e \right>) \beta^R  \right]  \; , \label{eq2:equation_of_motion_interacting_simplified_reservoir}
\end{equation}
where $\beta^R=\frac{1}{k_B T_R}$, $T_R$ is the temperature of the reservoir, and the superscripts, $A$, are removed because there is just one system of interest to follow. %(see Fig.\;\ref{fig2:system_reservoir}) 
%\begin{figure}
%\begin{center}
%\includegraphics[scale=0.42]{fig2_system_reservoir}
%\caption{\label{fig2:system_reservoir} The system interacting with a heat reservoir.   }
%\end{center}
%\end{figure}

Although only two subsystems exchanging energy in a heat interaction are considered here, the approach can be generalized to additional subsystems exchanging heat and/or mass \cite{li2016generalized}.

\section{\label{chap2_sec:level4}Pseudo-eigenstructure}
\begin{figure}
\begin{center}
\includegraphics[scale=0.37]{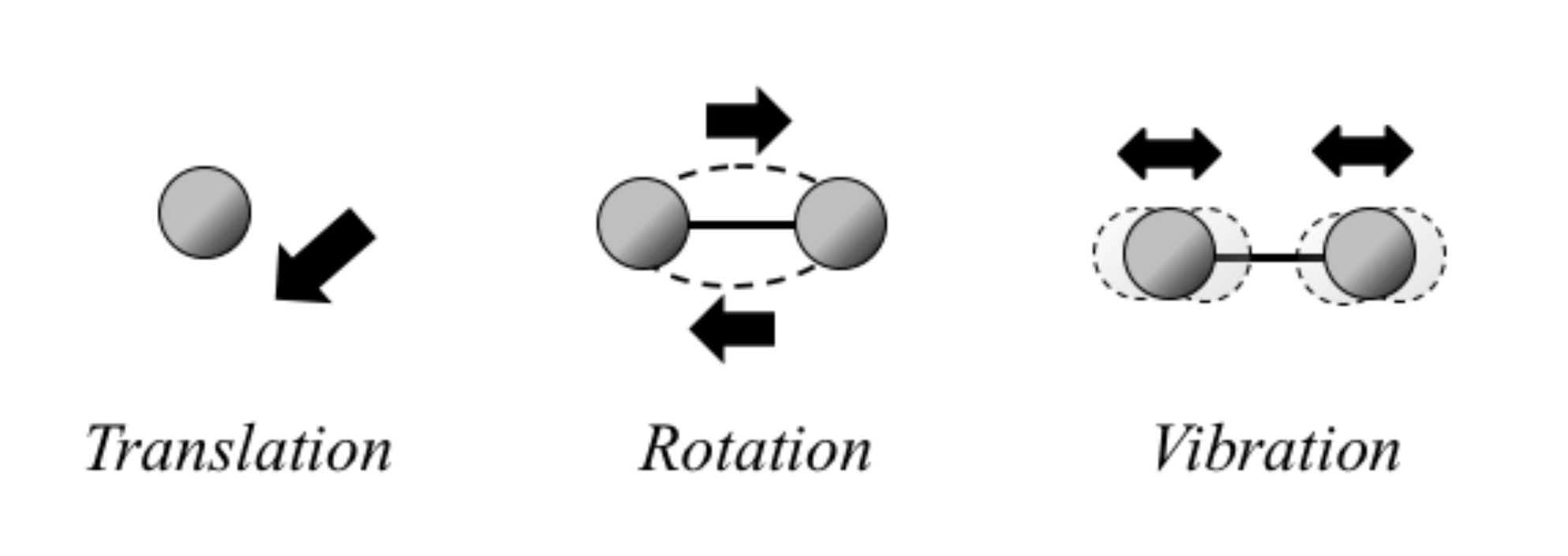}
\caption{\label{fig2:quantum_models} The translational, rotational, and vibrational degrees of freedom of particles (or molecules). They are commonly used as quantum models when an energy eigenstructure of a gas phase is constructed.  }
\end{center}
\end{figure}

The SEAQT equation of motion is solved with a particular energy eigenstructure. In general, an energy eigenstructure (a set of energy eigenlevels) is constructed for a quantum system by assuming appropriate degrees of freedom for the particles or molecules: for example, translation, rotation, and vibration degrees of freedom (see Fig.\;\ref{fig2:quantum_models}). A relatively simple energy eigenstructure can be constructed for a low-density gas by assuming the gas particles behave independently (the ideal gas approximation). In the solid (or liquid) phase, on the other hand, interactions between particles play a determining role for the properties so that interactions cannot be ignored and the energy eigenstructure becomes quite complex. This complexity can be mitigated by replacing the quantum model with a reduced-order model \cite{yamada2018method,yamada2018kineticpartI,yamada2018kineticpartII,yamada2018magnetization} constructed from an appropriate solid-state analog. Furthermore, since these energy eigenstructures usually involve an infinite number of energy eigenlevels --- and cannot be used with the SEAQT framework for this reason --- a density of states method \cite{li2016steepest} must be employed to convert an infinite energy-eiegnlevel system to a finite-level one. Two common reduced-order models (coupled oscillators and the mean-field approximation) are described in Sec.\;\ref{chap2_sec:level4_1} and the density of states method is explained in Sec.\;\ref{chap2_sec:level4_2}.

\subsection{\label{chap2_sec:level4_1}Reduced-order model}

\subsubsection{\label{chap2_sec:level4_1_1}Coupled oscillators}
\begin{figure}
\begin{center}
\includegraphics[scale=0.22]{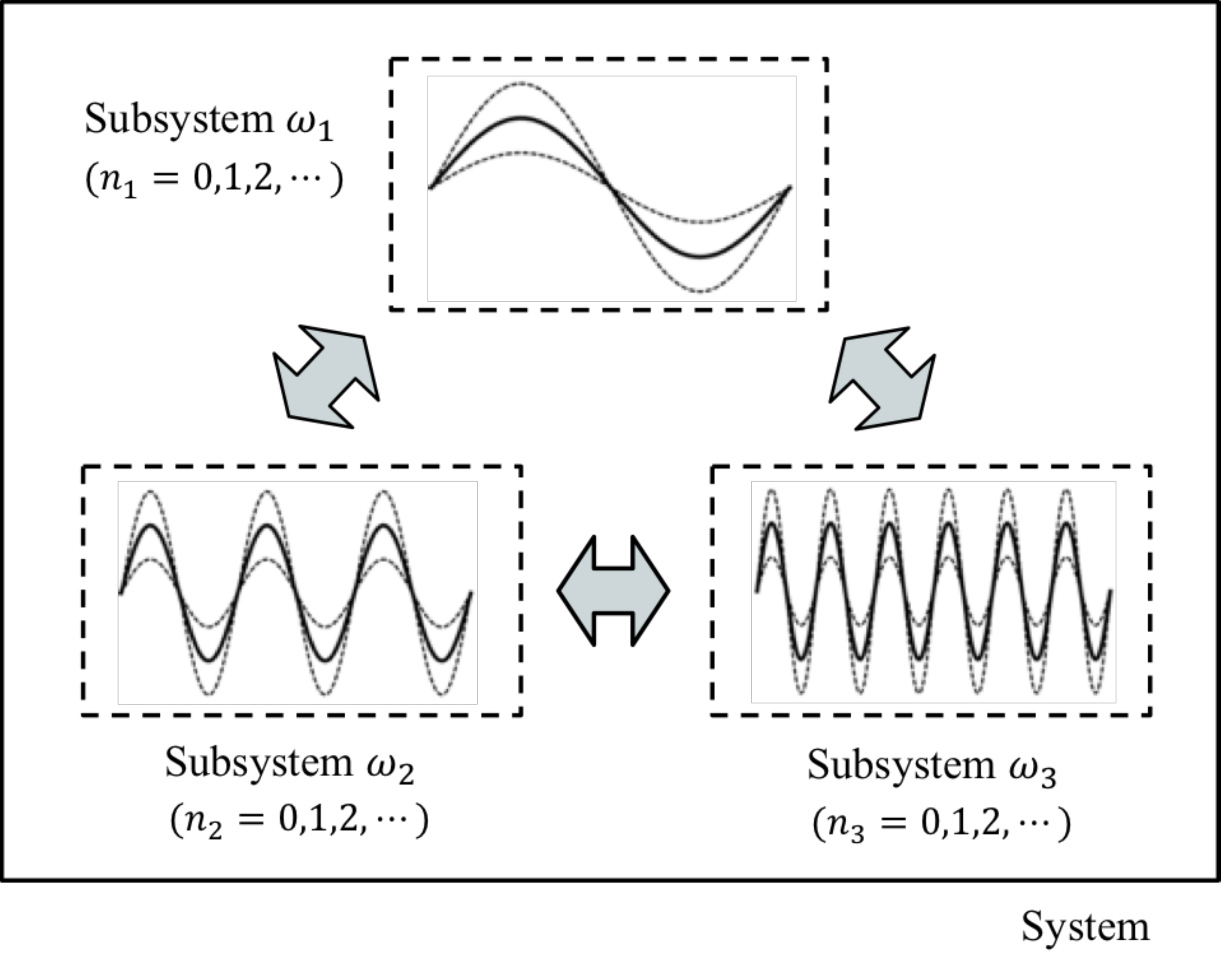}
\caption{\label{fig2:couple_multiple} The system description for coupled oscillators with various vibrational frequencies, $\omega_j$. The system is divided into three interacting subsystems, each with its own vibrational frequency. $n$ is an integer quantum number that applies to the phenomenon of interest, e.g., phonons for oscillating molecules in a lattice or magnons for magnetic spin on a lattice.   }
\end{center}
\end{figure}

Unlike atoms or molecules in a gas or liquid phase which include all of the degrees of freedom of Fig.\;\ref{fig2:quantum_models}, the motion of particles in a solid are spatially constrained and only include the vibrational degree of freedom. This has some computational benefits because it removes the need to calculate any eigenlevels associated with translation or rotation. Since atoms in a lattice exhibit collective atomic movements even at quite high temperatures, they can be modeled reasonably well by a collection of coupled oscillators with quantized energies.  The energy eigenstructure is constructed by associating energies with all the frequencies available to the system. This can be done by constructing a reduced-order model that treats a system of particle oscillators as a collection of subsystems with different vibrational frequencies (see Fig.\;\ref{fig2:couple_multiple}). The oscillators may be physical objects, like atoms or molecules, or they can be analogs like magnetic spin waves. Example applications of the approach are found in reference \cite{yamada2018method} where thermal expansion is calculated from an eigenstructure built from anharmonic coupled oscillators and in reference \cite{yamada2018magnetization} where magnetization is calculated from an eigenstructure based on harmonic coupled oscillators.

\subsubsection{\label{chap2_sec:level4_1_2}Mean-field approximation}
\begin{figure}
\begin{center}
\includegraphics[scale=0.37]{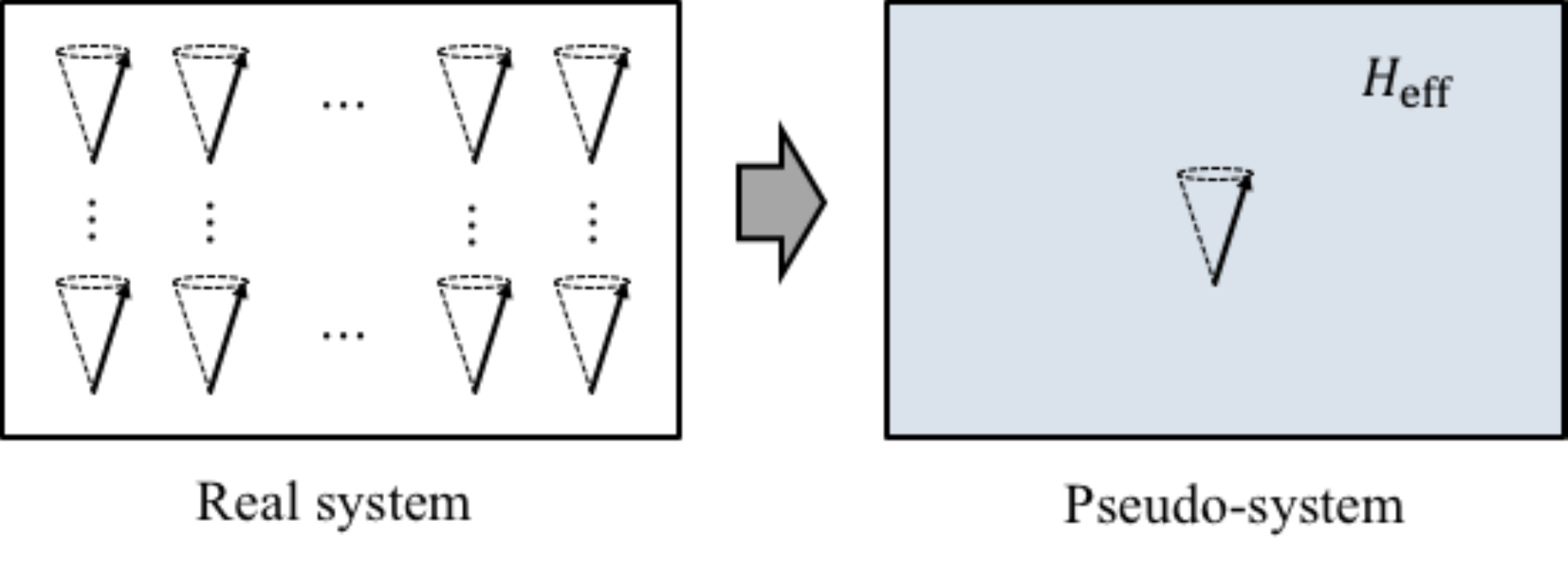}
\caption{\label{fig2:mean_field} The spin system before and after the mean-field approximation is employed. The interactions between magnetic moments (spins) is substituted by the effective internal magnetic field, $H_{\mbox{\scriptsize eff}}$. }
\end{center}
\end{figure}

The lattice (spin) wave description using coupled harmonic oscillators breaks down at high temperatures because of interactions among the subsystems of Fig.\;\ref{fig2:couple_multiple} (phonon-phonon or magnon-magnon interactions). These interactions can be included explicitly in the eigenstructure by using anharmonic oscillators rather than simple harmonic oscillators (see reference \cite{yamada2018method}). Alternatively, one can use a mean-field approximation to describe the interactions. The mean field approximation has been used extensively to describe the magnetization of ferromagnetic materials \cite{girifalco2003statistical,kittel1980thermal,aharoni2000introduction} where interactions among spins on a lattice are replaced with an effective internal magnetic field (see Fig.\;\ref{fig2:mean_field}). The mean-field model is often used with the Ising model where magnetic moments are allowed to point in only two directions, up or down. The method is illustrated in Sec.\;\ref{chap2_sec:level5_2} wherein the magnetization change of body-centered cubic (bcc) iron is calculated with the SEAQT framework.

\begin{figure}
\begin{center}
\includegraphics[scale=0.37]{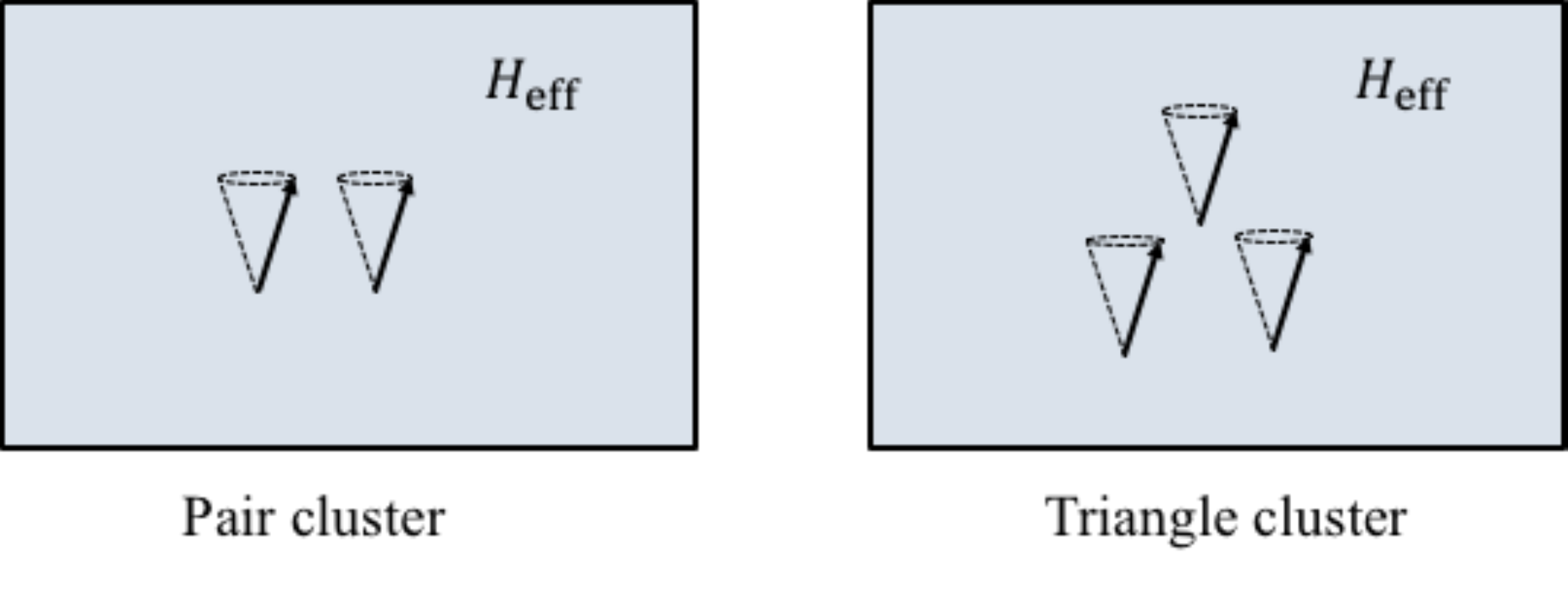}
\caption{\label{fig2:mean_field_clusters} The mean-field approximation, which includes short-range correlations by defining pair and triangle clusters, respectively. }
\end{center}
\end{figure}

The mean-field approximation fails to predict magnetization changes of ferromagnetic materials at low temperatures because it uses a uniform (or constant) value for the effective internal field and ignores changes of the field in the region where up-spins or down-spins are slightly localized. This happens at low temperatures because the contribution of interaction energy becomes large. To cope with the problem, there have been attempts to include short-range correlations between spins in the model by defining clusters \cite{girifalco2003statistical} (see Fig.\;\ref{fig2:mean_field_clusters}). The same is true for mean-field approximations applied to atomic configurations in alloys (see below) \cite{kikuchi1951theory}. However, very large clusters are required to describe the wave-like behavior of magnetic moments at low temperatures so the mean-field approximation is not suitable for describing magnetization at very low temperatures.  

Combining the mean-field approximation with an Ising model can also be used  to model atomic configurations in a binary $A$--$B$ alloy \cite{girifalco2003statistical,kittel1980thermal} where up- and down-spins are used to represent $A$- and $B$-atoms. The mean-field approximation replaces detailed interaction energies between particles with an effective interaction energy (as is done in a spin system, Fig.\;\ref{fig2:mean_field}). In this case, applying the SEAQT equation of motion to the eigenstructure can track the time-evolution of atomic arrangements in a specific alloy provided the atomic configurations are constrained to reflect the accessible states of the system as it evolves. This methodology is used in references \cite{yamada2018kineticpartI,yamada2018kineticpartII} to explore phase decomposition in a binary alloy system.

\subsection{\label{chap2_sec:level4_2}Density of states method}
As the number of oscillators or particles in a solid phase increases, the number of energy eigenlevels becomes effectively infinite, and applying the SEAQT equation of motion results in a system of equations infinite in extent, which clearly is problematic. This difficulty can be avoided with the density of states method developed by Li and von Spakovsky within the SEAQT framework \cite{li2016steepest}. The density of states method approximates an infinite energy-eigenlevel system with one composed of a finite number of discretized energy eigenlevels called a pseudo-eigenstructure. The approach is based on the observation that the occupation probabilities for all eigenlevels within a sufficiently small energy range behave dynamically in a similar fashion. As a result, the energy eigenlevels within a given range can be represented by a single pseudo-eigenlevel and associated degeneracy. A quasi-continuous condition \cite{li2016steepest} on the size of the energy range ensures that the approximate pseudo-eigenstructure effectively results in the same property values as would be predicted with the original infinite-level eigenstructure.

In the density of states method, the continuous energy distribution, $\epsilon (x)$, of an infinite-level energy system is divided into discrete bins with a set of discrete eigenlevels, $\epsilon_j$. The system with the continuous distribution of energy eigenlevels is referred to as the `original' system, and the discretized bins and associated energy eigenlevels as the `pseudo-system'. From a practical standpoint, the number of bins, $R$, in the pseudo-system is made as small as possible to reduce the number of simultaneous equations of motion that need to be solved in the SEAQT framework. However, in order to accurately represent the original energy eigenstructure, the property values predicted for the original and the pseudo-systems should be approximately same. The conditions under which this will be true can be established using canonical distributions. For the original system with a continuous energy spectrum, the canonical distribution for occupation probabilities is given by
\begin{equation}
p(x)=\frac{ g(x) e^{ - \beta \epsilon(x) }}{\int_{-\infty}^{\infty} g(x') e^{ - \beta \epsilon(x') } \; dx'}=\frac{ g(x) e^{ -\beta \epsilon(x)} }{Z^{\mathrm{cont}}} \; ,  \label{eq2:continuous-canonical_distribution}
\end{equation}
where $p(x)$ and $g(x)$ are, respectively, the occupation probability and the degeneracy of the energy $\epsilon(x)$ and $\beta=1/k_BT$. The occupation probability of a discrete energy eigenlevel in the pseudo-system, $p_j$, is expressed as 
\begin{equation}
\begin{split}
 p_j & = \int_{x^{\mathrm{min}}_j}^{x^{\mathrm{max}}_j} p(x) \; dx = \frac{1}{Z^{\mathrm{cont}}} \int_{x^{\mathrm{min}}_j}^{x^{\mathrm{max}}_j} g(x) e^{ - \beta \epsilon(x) } \; dx  \\
& = \frac{Z}{Z^{\mathrm{cont}}} \frac{1}{Z} e^{- \beta \epsilon_j} \int_{x^{\mathrm{min}}_j}^{x^{\mathrm{max}}_j} g(x) e^{ - \beta (\epsilon(x) - \epsilon_j) } \; dx  \; , 
\end{split}
\label{eq2:quasi_continuous_condition_derivation}
\end{equation}
where $x^{\mbox{\footnotesize{min\,(or max)}}}_j$ is the minimum (maximum) value of $x$ in the $j^{th}$ energy interval (or bin) and $\epsilon_j$ and $Z$ are, respectively, the $j^{th}$ energy eigenlevel and the partition function in the pseudo-system. When $$\frac{Z}{Z^{\mathrm{cont}}} \; e^{- \beta \left(\epsilon(x) - \epsilon_j \right)}  \approx 1$$ in the range, $x^{\mbox{\footnotesize{min}}}_j \leq x \leq x^{\mbox{\footnotesize{max}}}_j $, Eq.\;(\ref{eq2:quasi_continuous_condition_derivation}) can be written as
\begin{equation}
\begin{split}
 p_j & \approx \frac{1}{Z} e^{- \beta \epsilon_j} \int_{x^{\mathrm{min}}_j}^{x^{\mathrm{max}}_j} g(x) \; dx = \frac{g_j e^{-\beta \epsilon_j}}{Z}  \; , 
\end{split}
\label{eq2:quasi_continuous_condition_derivation2}
\end{equation}
where $g_j = \int_{x^{\mathrm{min}}_j}^{x^{\mathrm{max}}_j} g(x) \; dx$. Since Eq.\;(\ref{eq2:quasi_continuous_condition_derivation2}) is the canonical distribution for discrete energy eigenlevels, the property values of the original and pseudo-systems will be similar when the following condition is satisfied:
\begin{equation}
\begin{split}
 \frac{ \epsilon(x)-\epsilon_j }{k_BT } \approx \mathrm{ln}\left( \frac{Z}{Z^{\mathrm{cont}}} \right) =\mathrm{ln}\left( \frac{\sum_i g_i e^{-\beta \epsilon_i} }{\int_{-\infty}^{\infty} g(x') e^{ - \beta \epsilon(x') } \; dx'} \right) \; . 
\end{split}
\label{eq2:quasi_continuous_condition}
\end{equation}
When $Z \approx Z^{\mathrm{cont}}$, the condition can be simplified to 
\begin{equation}
\begin{split}
\frac{ \epsilon(x)-\epsilon_j }{k_BT } &\approx 0 \;  \Rightarrow \;  |\epsilon(x)-\epsilon_j|  \ll k_BT  \; \\ 
\Rightarrow \;&\; |\epsilon_{j \pm 1}-\epsilon_j|  \ll k_BT  \; , 
\end{split}
\label{eq2:quasi_continuous_condition_simplified}
\end{equation}
where the relation, $|\epsilon_{j \pm 1}-\epsilon_j| < |\epsilon(x)-\epsilon_j|$, is employed since $\epsilon_{j-1} < \epsilon(x^{\mathrm{min}}_j ) < \epsilon(x) < \epsilon(x^{\mathrm{max}}_j ) < \epsilon_{j+1} $ for a monotonic function of $\epsilon(x)$. Thus, when $Z \approx Z^{\mathrm{cont}}$, the number of energy intervals (or bins), $R$, can be determined by checking whether Eq.\;(\ref{eq2:quasi_continuous_condition_simplified}), which is called the quasi-continuous condition \cite{li2016steepest}, is satisfied or not. Note that since $Z < Z^{\mathrm{cont}}$ in most cases, the general condition, Eq.\;(\ref{eq2:quasi_continuous_condition}), is less stringent than that given by Eq.\;(\ref{eq2:quasi_continuous_condition_simplified}).

\section{\label{chap2_sec:level5}Demonstrations}

\subsection{\label{chap2_sec:level5_1}Simple model systems}
The use of the SEAQT equation of motion is illustrated in this section assuming a simple system composed of particles with four, non-degenerate energy eigenlevels. This model was introduced in reference \cite{beretta2006nonlinear} for an isolated system. Here, interactions with a heat reservoir or another system are considered. 

\begin{figure}
\begin{center}
\includegraphics[scale=0.53]{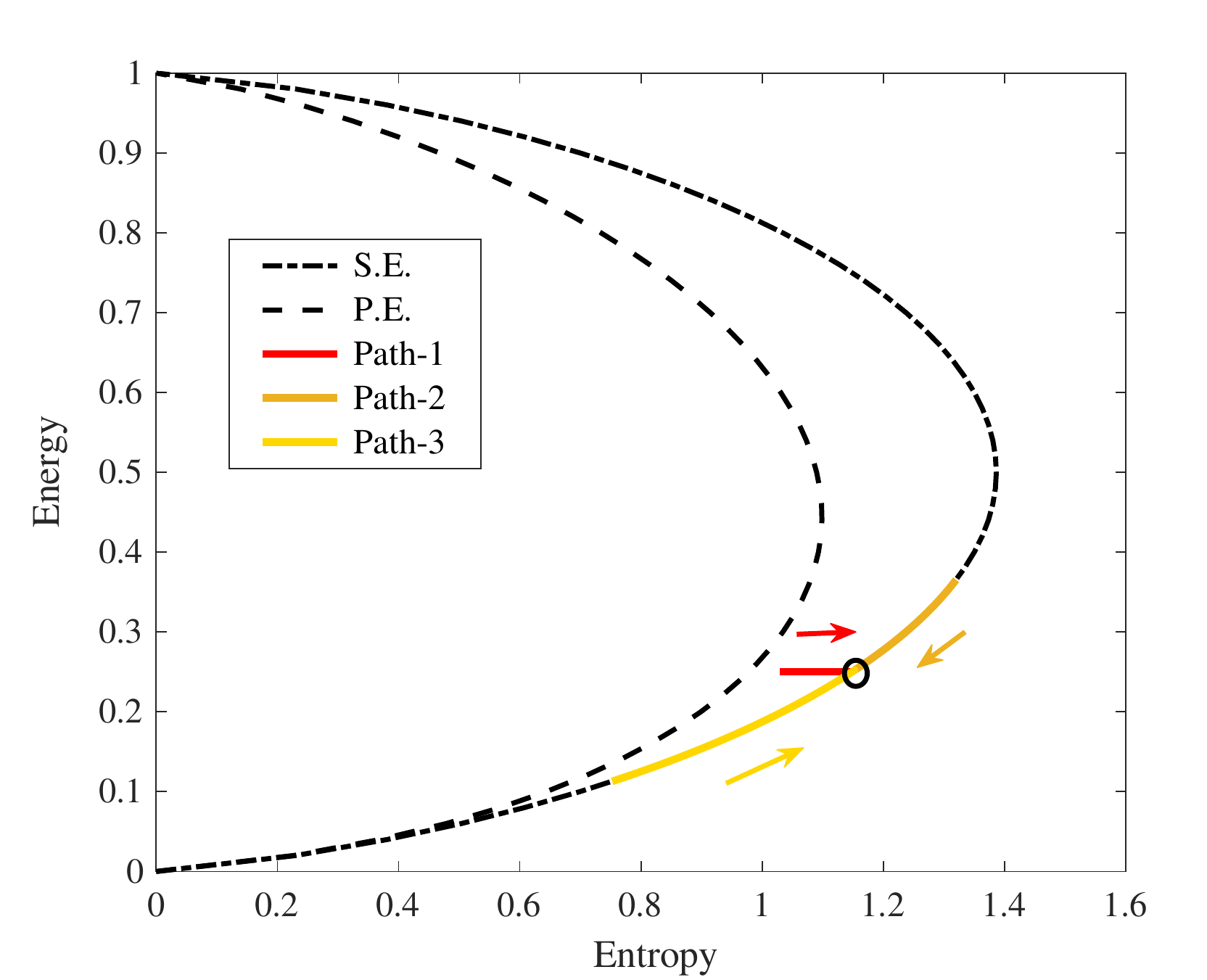}
\caption{\label{fig2:E_S_diagram_three} The \textit{E}--\textit{S} diagram for a system of particles with four energy eiegenlevels, $[\epsilon_1,\epsilon_2,\epsilon_3,\epsilon_4]=[0, 1/3, 2/3, 1]$. \cite{beretta2006nonlinear} The entropy and energy calculated from the canonical (stable equilibrium) and partially canonical distributions are represented by the dotted and broken lines, respectively. Three different kinetic paths calculated using the SEAQT equation of motion are labeled with arrows. Path-1 is for an isolated system whose initial non-equilibrium state is prepared by Eq.\;(\ref{eq2:initial_probability_partial}) with $\lambda=0.1$. Paths 2 and 3 are for a system interacting with a heat reservoir, $T_R$, evolving from different initial states prepared using Eq.\;(\ref{eq2:canonical_distribution}): Path-2 represents cooling from $T_0=1.0$, and Path-3 depicts heating from $T_0=0.25$. The final stable equilibrium state for all three paths is indicated by the open circle and corresponds to a temperature of $T^{\mbox{\footnotesize se}}$ or $T_R = 0.5$.  }
\end{center}
\end{figure}

The four energy eiegenlevels, $\epsilon_j$, are arbitrarily set as $[\epsilon_1,\epsilon_2,\epsilon_3,\epsilon_4]=[0, 1/3, 2/3, 1]$ with no degeneracy, i.e., the $g_j=1$. The stable equilibrium states can be determined by the canonical distribution:
\begin{equation}
p^{\mbox{\footnotesize{se}}}_j=\frac{g_j \mathrm{exp}( -\beta^{\mbox{\footnotesize{se}}} \epsilon_j )}{\sum\limits_i g_i \mathrm{exp}( -\beta^{\mbox{\footnotesize{se}}} \epsilon_i )}=\frac{ g_j \mathrm{exp}(-\beta^{\mbox{\footnotesize{se}}} \epsilon_j )}{Z^{\mbox{\footnotesize{se}}}} \; ,  \label{eq2:canonical_distribution}
\end{equation}
where $Z^{\mbox{\footnotesize{se}}}$ is the partition function, $\beta^{\mbox{\footnotesize{se}}}=1/k_BT^{\mbox{\footnotesize{se}}}$, and the $\mbox{\small{se}}$ superscript denotes stable equilibrium. 
Consider now a system in which some of the available energy eigenlevels are not occupied; such a system is not in stable equilibrium. The occupation probabilities calculated with a canonical distribution modified to account for the unoccupied energy eigenlevels are referred to a partially canonical distribution \cite{beretta2006nonlinear}: 
\begin{equation}
p^{\mbox{\footnotesize{pe}}}_j= \frac{\delta_j g_j \mathrm{exp}( -\beta^{\mbox{\footnotesize{pe}}} \epsilon_j )}{\sum\limits_i \delta_i g_i \mathrm{exp}( -\beta^{\mbox{\footnotesize{pe}}} \epsilon_i )} \; ,  \label{eq2:partial_canonical_distribution}
\end{equation}
where $\beta^{\mbox{\footnotesize{pe}}}=1/k_BT^{\mbox{\footnotesize{pe}}}$ and $\delta_j$ takes a value of one or zero depending upon whether the state is occupied or not. For four energy eiegenlevels, one could make a partially canonical distribution, for example, by making the third energy eigenlevel unoccupied, or setting $[\delta_1,\delta_2,\delta_3,\delta_4]=[1, 1, 0, 1]$ in Eq.\;(\ref{eq2:partial_canonical_distribution}). This partially canonical distribution can be used to determine an initial non-equilibrium state for the SEAQT equation of motion. The \textit{E}--\textit{S} diagram calculated from the canonical distribution, Eq.\;(\ref{eq2:canonical_distribution}), and the partially canonical distribution, Eq.\;(\ref{eq2:partial_canonical_distribution}), is shown in Fig.\;\ref{fig2:E_S_diagram_three}.   For simplicity in this illustrative example, dimensionless energies with $k_B=1$ are used.
% (if one is confused the dimensionless energy descriptions, he/she can think that they are normalized as $[\epsilon_1,\epsilon_2,\epsilon_3,\epsilon_4]/k_BT_{\mathrm{273K}}=[0, 1/3, 2/3, 1]$, where $T_{\mathrm{273K}}$ is a room temperature).

\begin{figure}
\begin{center}
\includegraphics[scale=0.53]{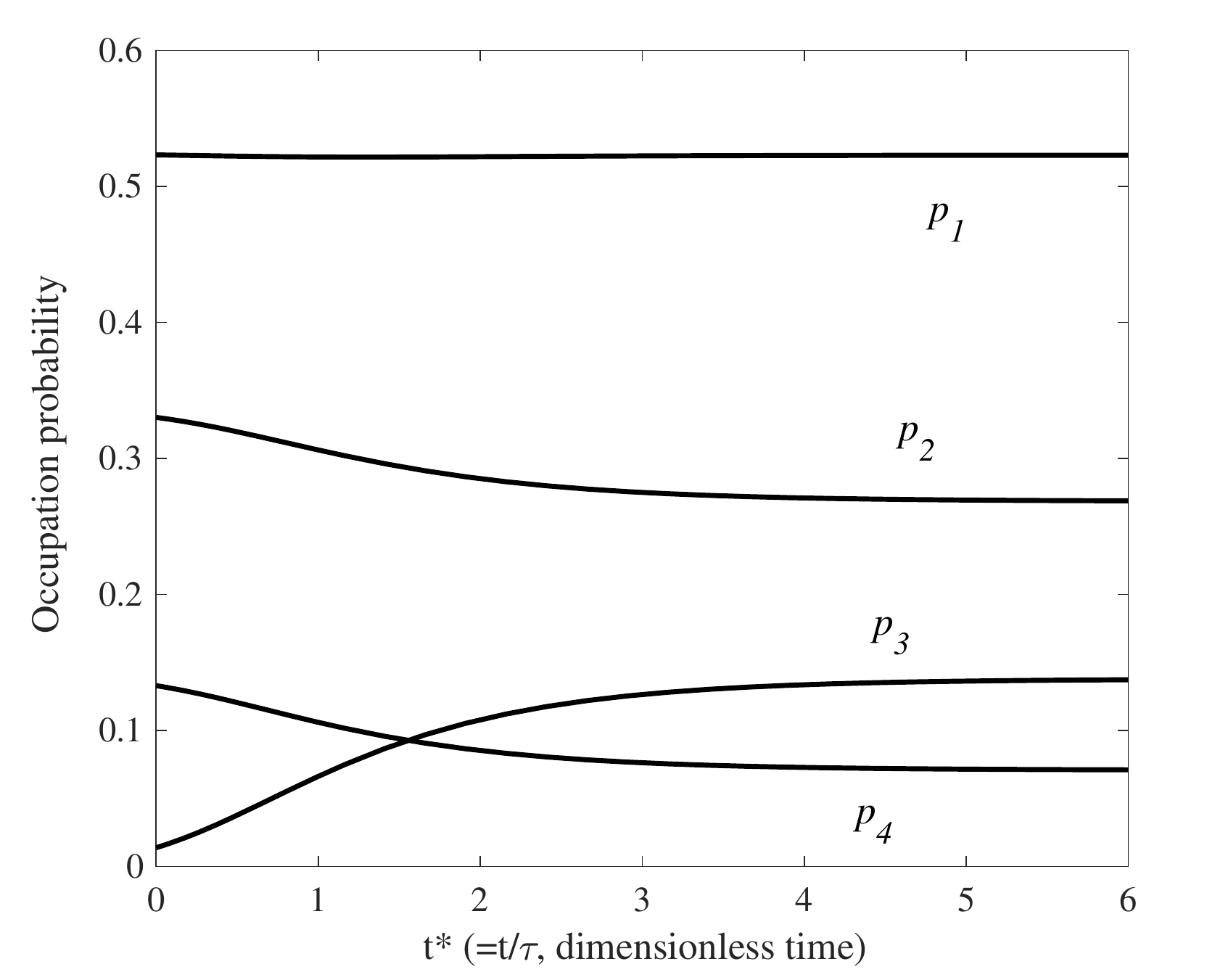}
\caption{\label{fig2:occupation_probability_isolated_pe} The time dependence of the occupation probabilities for the isolated system in the relaxation process shown in Fig.\;\ref{fig2:E_S_diagram_three} (Path-1). The detail analysis of the relaxation process for the isolated system can be found in reference \cite{beretta2006nonlinear}.  }
\end{center}
\end{figure}

First, three different relaxation paths are investigated using the SEAQT equation of motion. Two paths (Paths 2 and 3) represent a system moving from an initial equilibrium state to a final equilibrium state through an interaction with a heat reservoir, $T_R$, using Eq.\;(\ref{eq2:equation_of_motion_interacting_simplified_reservoir}). The initial states (or initial occupation probabilities), $p_j^0$, are prepared from Eq.\;(\ref{eq2:canonical_distribution}) by replacing $T^{\mbox{\footnotesize{se}}}$ with $T_{0}$ of a chosen value for the initial temperature. The remaining path (Path-1) corresponds to an isolated system evolving from a non-equilibrium initial state to stable equilibrium using Eq.\;(\ref{eq2:equation_of_motion_simplified}). The initial state for this path is given using the partially canonical distribution, $p^{\mbox{\footnotesize{pe}}}_j$, and a perturbation equation that displaces the initial state from the partially canonical state such that \cite{beretta2006nonlinear}
\begin{equation}
p^0_{j}=(1-\lambda_{\mbox{\scriptsize const}})p^{\mbox{\footnotesize pe}}_{j}+\lambda_{\mbox{\scriptsize const}}p^{\mbox{\footnotesize se}}_{j} \; ,  \label{eq2:initial_probability_partial}
\end{equation}
where $\lambda_{\mbox{\scriptsize const}}$ is the perturbation constant. Note that $\beta^{\mbox{\footnotesize{pe}}}$ is determined through the relation, $\sum_{i} \epsilon_{i} p^{\mbox{\footnotesize pe}}_{i} =\sum_{i} \epsilon_{i} p^{\mbox{\footnotesize se}}_{i} $. The calculated kinetic paths are shown in Fig.\;\ref{fig2:E_S_diagram_three} as well as the canonical and partially canonical distributions. As can be seen, although the initial states are different, the final states are the same and correspond to a stable equilibrium state at $T_R$ (or $T^{\mbox{\footnotesize se}}$). The time-dependence of each occupation probability in the relaxation process for the isolated system (Path-1 of Fig.\;\ref{fig2:E_S_diagram_three}) is shown in Fig.\;\ref{fig2:occupation_probability_isolated_pe}. Although the expected energy, $\left< e \right>=\sum_i \epsilon_i p_i$, is constant throughout the process (Path-1 is a horizontal line on the \textit{E}--\textit{S} diagram of Fig.\;\ref{fig2:E_S_diagram_three}), the probability distribution among the individual energy eigenlevels does change with time. This {\em redistribution} of the internal energy is driven by an increase in entropy as the state of the system evolves.

\begin{figure}
\begin{center}
\includegraphics[scale=0.52]{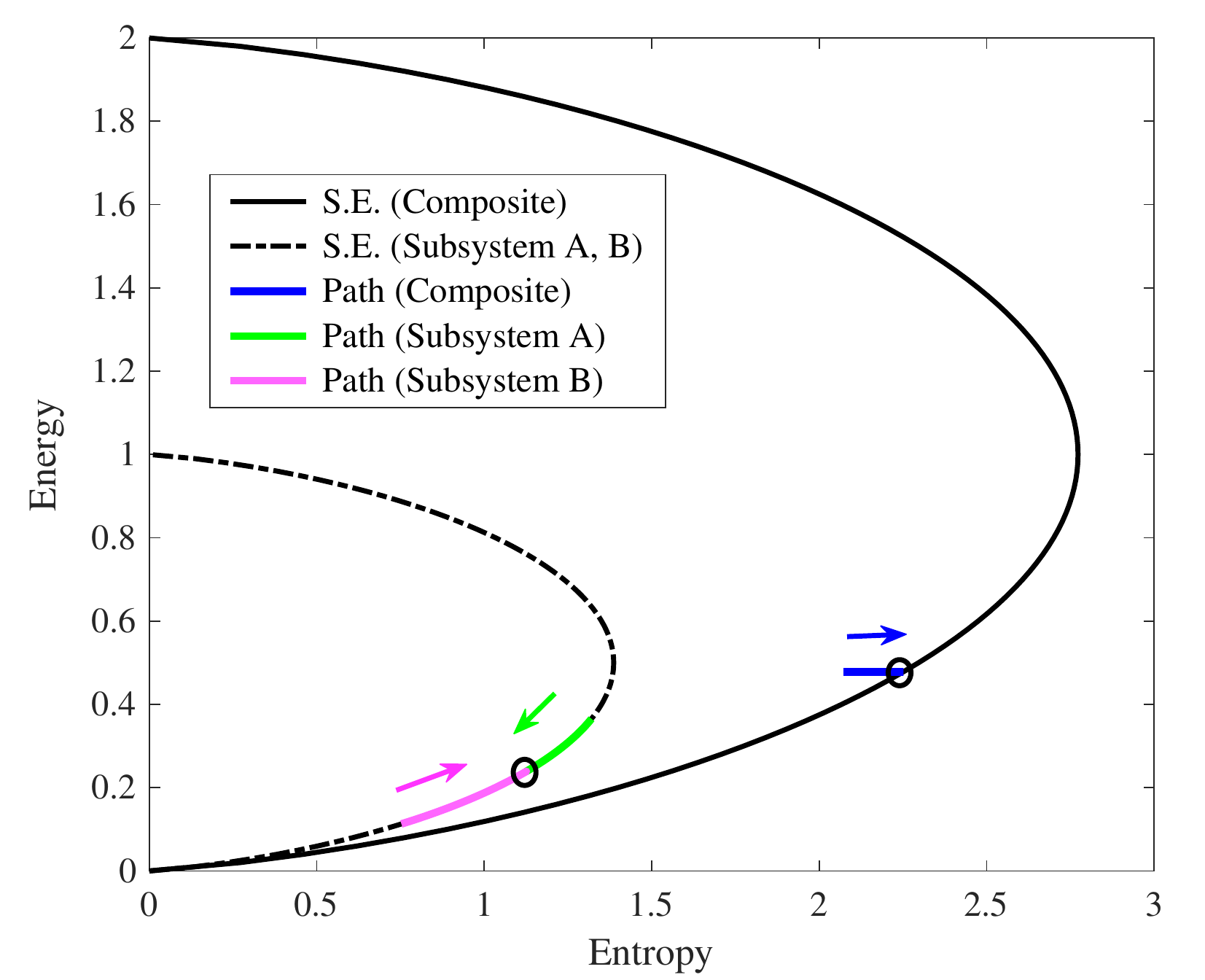}
\caption{\label{fig2:E_S_diagram_composite} The \textit{E}--\textit{S} diagrams for the interacting subsystems\;$A$ and $B$ as well as for the composite system, $A+B$. The canonical distributions of the composite system and subsystems\;$A$ and $B$ are, respectively, shown as solid and dotted lines. The kinetic paths of each subsystem and the composite calculated using the SEAQT equation of motion are depicted as well. The final states of the subsystems and the composite are shown by open circles. }
\end{center}
\end{figure}

\begin{figure}
\begin{center}
\includegraphics[scale=0.57]{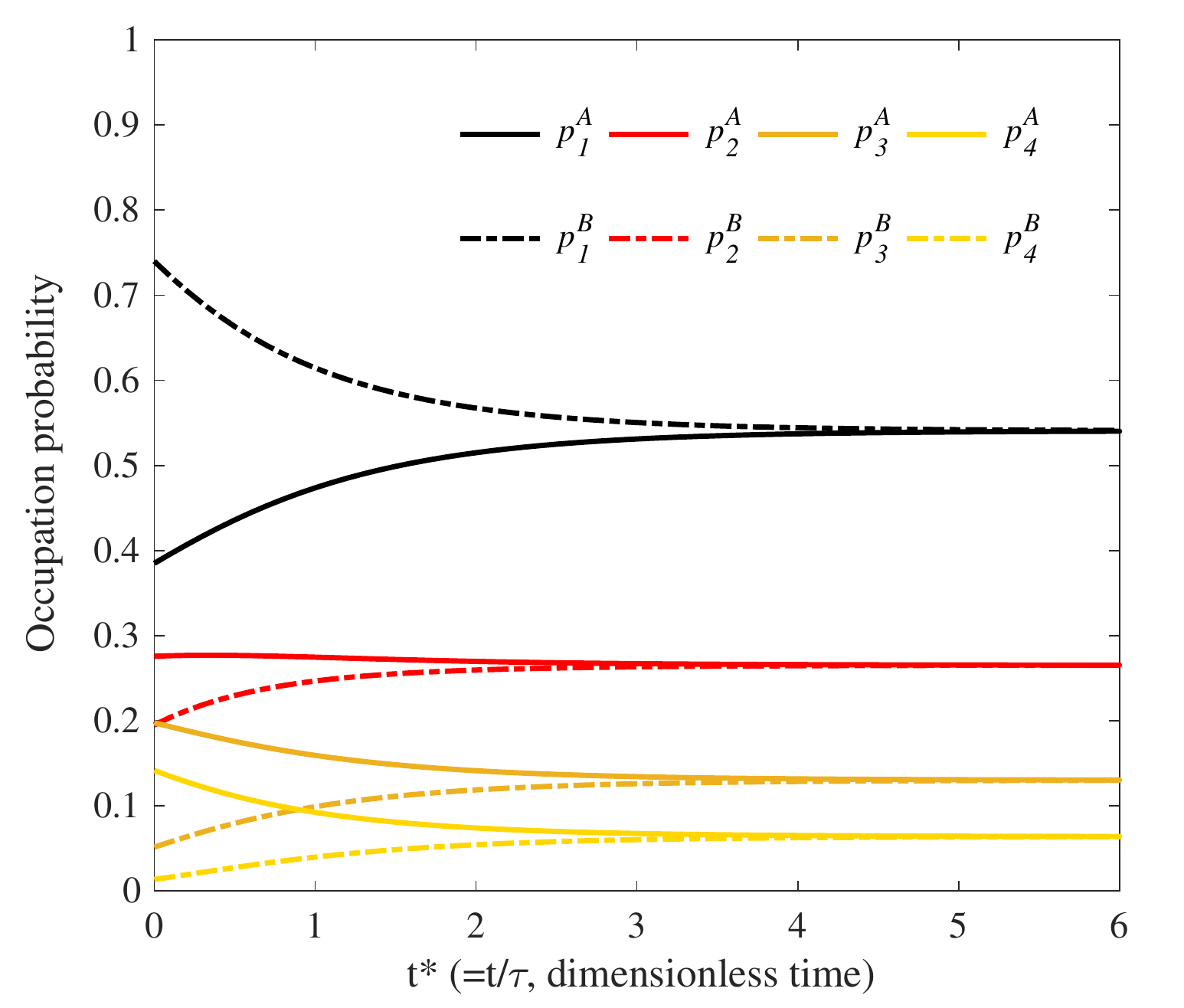}
\caption{\label{fig2:occupation_probability_AandB} The time dependence of occupation probabilities in subsystems\;$A$ and $B$ in the relaxation process shown in Fig.\;\ref{fig2:E_S_diagram_composite}. The occupation probabilities of subsystems\;$A$ and $B$ are shown by solid and dotted lines, respectively. }
\end{center}
\end{figure}

Next, a case involving a heat interaction between two systems is considered. The two systems are treated within the context of an isolated composite system. This correspond to the system description of Fig.\;\ref{fig2:isolated_systems}\;(b). The two subsystems, $A$ and $B$, are identical and have the same energy eigenstructure as described above. The \textit{E}--\textit{S} diagrams calculated for subsystems\;$A$ and $B$ as well as the composite system, $A+B$, using Eq.\;(\ref{eq2:canonical_distribution}) are shown in Fig.\;\ref{fig2:E_S_diagram_composite}. Because both the energy and entropy are extensive properties, the values of these properties for the composite system are twice the energy and entropy of the individual subsystems\;$A$ and $B$. The relaxation paths of each subsystem calculated from Eq.\;(\ref{eq2:equation_of_motion_interacting}) (or Eq.\;(\ref{eq2:equation_of_motion_interacting_simplified})) are shown together in Fig.\;\ref{fig2:E_S_diagram_composite} where initial states are prepared by Eq.\;(\ref{eq2:canonical_distribution}) with $T^A_0=1.0$ and $T^B_0=0.25$. While the energy in the composite system is constant, the energies of subsystems\;$A$ and $B$ are not, and they approach each other with time and reach the same final states, which indicates they are in a mutual stable equilibrium (i.e., $T^A=T^B$). The time evolution of the occupation probabilities in subsystems\;$A$ and $B$ are shown in Fig.\;\ref{fig2:occupation_probability_AandB}. Although the initial probability distributions are different in the two subsystems, they become the same at the final state of mutual stable equilibrium. Recall that the two subsystems here are assumed to be identical. If they are not, the probability distributions are not necessarily the same even at mutual stable equilibrium.

Note that as can be seen in Fig.\;\ref{fig2:E_S_diagram_composite}, the kinetic path of each subsystem moves along its own manifold of different stable equilibrium states. This is a direct result of the steepest-entropy-ascent principle when initial states belong to the manifold and is an essential feature of the concept of hypo-equilibrium states \cite{li2016steepest,li2016generalized} described in Appendix\;\ref{chap2_sec:level7_2}. The non-equilibrium state of the composite system, $A+B$, at every instant of time is what Li and von Spakovsky call a $2^{nd}$-order hypo-equilibrium state.

\subsection{\label{chap2_sec:level5_2}bcc-Fe spin system}
To extend beyond a simple system model, a realistic magnetic spin system is considered next and a pseudo-eigenstructure is constructed based on a reduced-order model (an Ising model with a mean-field approximation) and the density of states method. The SEAQT equation of motion is applied to the pseudo-system eigenstructure to calculate the magnetization of bcc-Fe in the presence of an external magnetic filed and a heat reservoir.

\subsubsection{\label{chap2_sec:level5_2_1}Theory}
The SEAQT equation of motion for a ferromagnetic material is derived first. When magnetic spin is conserved, the manifold is $L(\bm{g}_I, \bm{g}_E, \bm{g}_M)$ (where $\bm{g}_M$ is the gradient of the magnetization) and the SEAQT equation of motion becomes   
\begin{equation}
\frac{dp_j}{dt^*}=\frac{\begin{vmatrix} 
-p_j \mathrm{ln} \frac{p_j}{g_j} & p_j & \epsilon_j p_j  & m_j p_j  \\
\left< s \right> & 1 & \left< e \right> & \left< m \right> \\
\left< es \right> & \left< e \right> & \left< e^2 \right> & \left< em \right> \\
\left< ms \right> & \left< m \right> & \left< em \right> & \left< m^2 \right> \\
\end{vmatrix}}{\begin{vmatrix} 
1 &  \left< e \right> & \left< m \right> \\
 \left< e \right> & \left< e^2 \right> & \left< em \right> \\
 \left< m \right> & \left< em \right> & \left< m^2 \right> \\
\end{vmatrix}} \; ,  \label{eq2:equation_of_motion_magnetization}
\end{equation}
where 
\[
\begin{array}{c c}
\langle s \rangle = - \sum\limits_{i} p_i \mathrm{ln} \frac{p_i}{g_i}  \; ,
&
\langle e \rangle = \sum\limits_{i} \epsilon_i p_i  \; , \\ \\
\langle m \rangle = \sum\limits_{i} m_i p_i  \; , \;\;\;
&
\langle es \rangle = - \sum\limits_{i} \epsilon_i p_i \mathrm{ln} \frac{p_i}{g_i}  \; , \\ \\
\langle e^2 \rangle = \sum\limits_{i} \epsilon_i^2 p_i   \; ,
&
\langle em \rangle = \sum\limits_{i} \epsilon_i m_i p_i  \; , \\ \\
\langle ms \rangle = - \sum\limits_{i} m_i p_i \mathrm{ln} \frac{p_i}{g_i}  \; , \;\;\;
&
\langle m^2 \rangle = \sum\limits_{i} m_i^2 p_i  \; ,
\end{array}
\]
and $m_j$ is the magnetization associated with the $j^{th}$ energy eigenlevel, $\epsilon_j$. When there is an exchange of energy via a heat interaction between the system of interest and a heat reservoir, $T_R$, in an external magnetic field, $H_R$, Eq.\;(\ref{eq2:equation_of_motion_magnetization}) is transformed into \cite{yamada2018magnetization}
\begin{equation}
\frac{dp_j}{dt^*}= p_j \left[ \left( s_j - \langle s \rangle \right)  - \left( \epsilon_j - \left< e \right> \right) \beta^R + \left( m_j- \langle m \rangle \right) \gamma^R  \right]  ,     
\label{eq2:equation_motion_magnetization_heat}
\end{equation}
where $\beta^R=1/k_BT_R$ and $\gamma^R=H_R/k_BT_R$.  

Next, a simplified eigenstructure is constructed using the Ising model and the mean-field approximation. When interactions between only the first-nearest-neighbor pairs are taken into account, the energy of the spin system is given by
\begin{equation}
E= \frac{1}{2} Nz \sum_{ij} e_{ij} \, y_{ij} \; ,     \label{eq2:energy_pair_interaction}
\end{equation}
where $N$ is the number of lattice points, $z$ is the coordination number (the number of first-nearest-neighbor sites per lattice point), and $e_{ij}$ and $y_{ij}$ are, respectively, the pair interaction energy and the pair (cluster) probability between $i$ and $j$ spins. When the mean-field approximation (with no short-range correlations) is employed, Eq.\;(\ref{eq2:energy_pair_interaction}) becomes (see Appendix\;\ref{chap2_sec:level7_3})
\begin{equation}
E(c)= \frac{1}{2} Nz \; J_{\mbox{\scriptsize eff}}\; c(1-c)  \; ,     \label{eq2:energy_pair_interaction_mean_field}
\end{equation}
where $c$ is the fraction of down-spins and $$J_{\mbox{\scriptsize eff}} \equiv 2e_{\uparrow \downarrow} - e_{\uparrow \uparrow} - e_{\downarrow \downarrow} \;.$$ The degeneracy of Eq.\;(\ref{eq2:energy_pair_interaction_mean_field}) is given by a binomial coefficient as
\begin{equation}
g(c)=\frac{N !}{N_{\uparrow}! N_{\downarrow} !}=\frac{N !}{(N(1-c))! (Nc)!} \; ,     \label{eq2:degeneracy_ising}
\end{equation}
where $N_{\uparrow}$ and $N_{\downarrow}$ are the number of lattice sites associated with up-spin and down-spin, respectively. Here, using the approximation for a factorial \cite{weisstein2008stirling}, $$x!\approx(2x+\frac{1}{3} \pi) x^x e^{-x}\;,$$ Eq.\;(\ref{eq2:degeneracy_ising}) is a continuous function. The energy eigenlevels and the degeneracy, $E_j$ and $g_j$, are determined from Eqs.\;(\ref{eq2:energy_pair_interaction_mean_field}) and (\ref{eq2:degeneracy_ising}) by replacing $c$ with $c_j$. In a bulk material, the atomic fraction of down-spin, $c_j$, could take any value, and the number of energy eigenlevels becomes effectively infinite. To cope with this infinity of levels, the density of states method \cite{li2016steepest} is used (see Sec.\;\ref{chap2_sec:level4_2}). Following the procedures of Sec.\;\ref{chap2_sec:level4_2}, the energy eigenlevels, degeneracies, and fractions of down-spins become 
\begin{equation}
E_j = \frac{1}{g_j} \int_{\bar{c}_j}^{\bar{c}_{j+1}} g(c') E (c') \; dc' \; ,\label{eq2:energy_eigenvalue_pseud_spin_flip}
\end{equation}
\begin{equation}
g_j=\int_{\bar{c}_j}^{\bar{c}_{j+1}} g(c') \; dc' \; ,  \label{eq2:degeneracy_pseud_spin_flip}
\end{equation}
and
\begin{equation}
c_j = \frac{1}{g_j} \int_{\bar{c}_j}^{\bar{c}_{j+1}}g(c')c' \; dc' \; ,\label{eq2:fraction_down_spin_pseud_spin_flip}
\end{equation}
where $\bar{c}_j$ is specified using the number of intervals, $R$, as $\bar{c}_j= j/R$. Here $j$ is an integer and takes values from zero to $R/2$. The magnetization for a given energy eigenlevel is given using the fraction of down-spins, $c_j$, as 
\begin{equation}
M_j = N \mu \left( 1-2 c_j  \right)  \; , \label{eq2:magnetization_pseud_spin_flip}
\end{equation}
where $\mu$ is the magnetic moment of iron ($\mu=2.22\mu_B$ where $\mu_B$ is the Bohr magneton \cite{kittel1966introduction}). Note that the energy eigenlevels and magnetizations are expressed here as $E_j$ and $M_j$ instead of $\epsilon_j$ and $m_j$ in order to emphasize that these are extensive properties.  

The number of intervals, $R$, is determined from Eq.\;(\ref{eq2:quasi_continuous_condition_simplified}) (or Eq.\;(\ref{eq2:quasi_continuous_condition})). However, since the degeneracy, $g_j$, in Eq.\;(\ref{eq2:degeneracy_pseud_spin_flip}) significantly increases with the number of particles, $N$, the ratio, $Z/Z^{\mathrm{cont}}$, rapidly decreases with $N$ and the criterion given in Eq.\;(\ref{eq2:quasi_continuous_condition_simplified}) becomes greatly relaxed. Taking this into account, the following relaxed criterion is used here instead of Eq.\;(\ref{eq2:quasi_continuous_condition_simplified}) (or Eq.\;(\ref{eq2:quasi_continuous_condition})):
\begin{equation}
 \frac{ | E_{j \pm 1}-E_j | }{N} \ll k_BT  \; .  \label{eq2:quasi_continuous_condition_extensive}
\end{equation}
The validity of Eq.\;(\ref{eq2:quasi_continuous_condition_extensive}) was tested for this particular application by repeating the calculations for different numbers of energy intervals to see how the calculated magnetization converges and then by confirming that the the results calculated based on Eq.\;(\ref{eq2:quasi_continuous_condition_extensive}) are close to the converged magnetization.

\subsubsection{\label{chap2_sec:level5_2_2}Results}
\begin{figure}
\begin{center}
\includegraphics[scale=0.5]{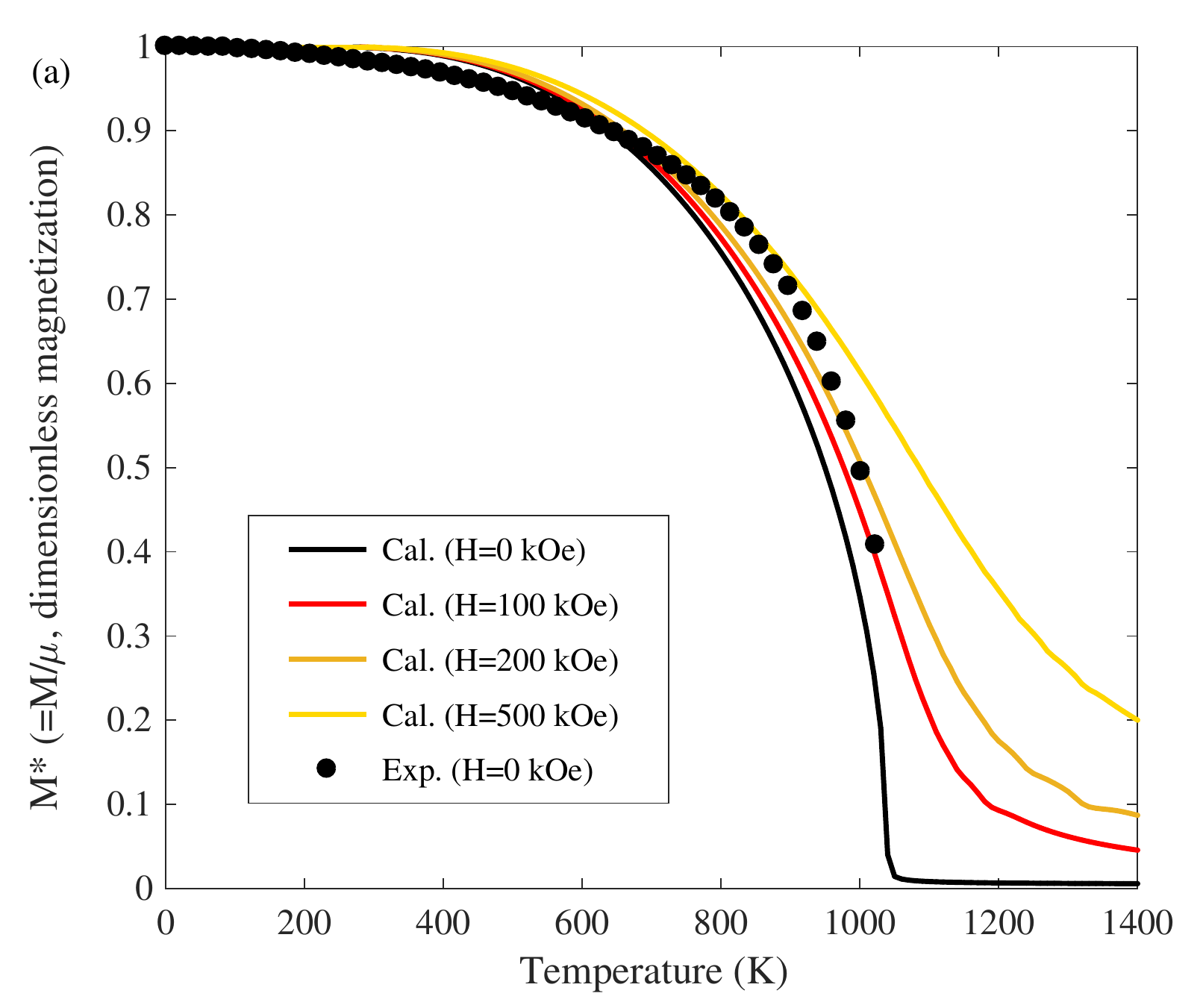}
\includegraphics[scale=0.5]{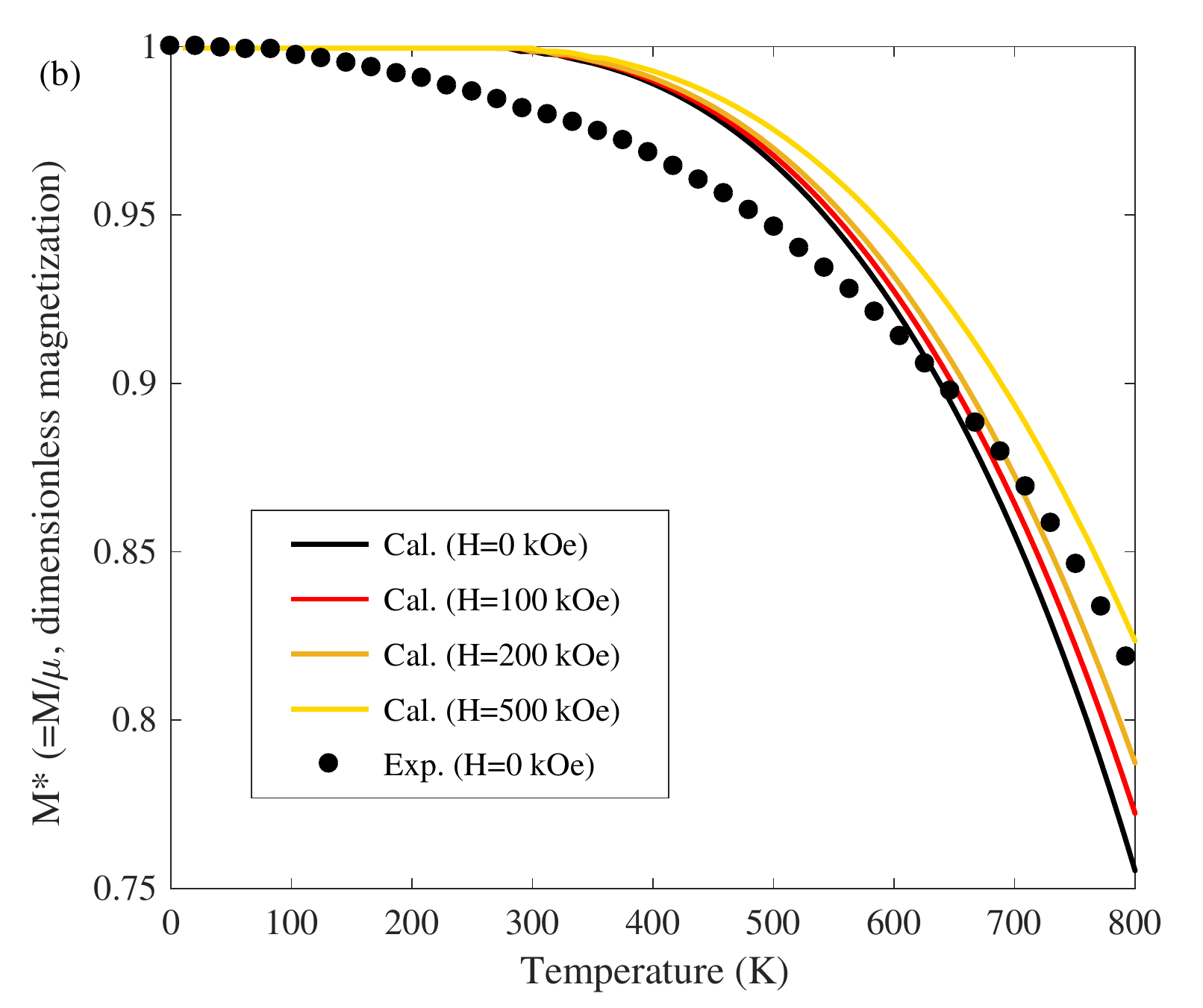}
\caption{\label{fig2:stable_magnetization} The calculated temperature dependence of equilibrium magnetizations of bcc-Fe at various external magnetic field strengths using $N=10^6$. (b) shows the low temperature range of (a). The solid black circles are experimental data at $H=0$ (kOe) \cite{crangle1971magnetization}. The magnetization, $M^*$, is a dimensionless magnetization normalized by the magnetic moment of iron, $M^*=M/\mu$. % $\mu=2.22\mu_B$,\cite{kittel1966introduction}
}
\end{center}
\end{figure}

The equilibrium magnetization at each temperature is determined from the extended canonical distribution
\begin{equation}
\begin{split}
p^{\mbox{\footnotesize se}}_j & =\frac{g_j \; \mathrm{exp}[ -\beta^{\mbox{\footnotesize se}} \left( E_j - M_j H^{\mbox{\footnotesize se}} \right) ]}{\sum\limits_i g_i \;\mathrm{exp}[ -\beta^{\mbox{\footnotesize se}} \left( E_i  - M_i H^{\mbox{\footnotesize se}} \right) ]} \\
& =\frac{g_j \; \mathrm{exp}[ -\beta^{\mbox{\footnotesize se}} \left( E_j - M_j H^{\mbox{\footnotesize se}} \right) ]}{Z^{\mbox{\footnotesize se}}} \; ,  \label{eq2:canonical_distribution_magnetization}
\end{split}
\end{equation}
where $Z^{\mbox{\footnotesize se}}$ is the partition function, $\beta^{\mbox{\footnotesize se}}$=1/$k_BT^{\mbox{\footnotesize se}}$, and $T^{\mbox{\footnotesize se}}$ and $H^{\mbox{\footnotesize se}}$ are, respectively, the temperature and the external magnetic field strength at stable equilibrium. The calculated temperature dependence of the magnetization, $M =  \sum_i M_i p_i$, in various external magnetic field strengths is shown in Fig.\;\ref{fig2:stable_magnetization} where $J_{\mbox{\scriptsize eff}}$ is estimated from experimental data of the Curie temperature, $T_c=1043$\;K, \cite{kittel1966introduction} as $J_{\mbox{\scriptsize eff}}=7.2 \times 10 ^{-22}$ (J/atom) (see Appendix\;\ref{chap2_sec:level7_3}). It can be seen that the calculated magnetization shows a similar temperature dependence with the experimental data and increases with an external magnetic field, as expected. However, the results at low temperatures deviate from experiments. This is a well-known tendency in the equilibrium magnetization calculated from the Ising model with a mean-field approximation \cite{girifalco2003statistical,kittel1980thermal,aharoni2000introduction} because spin wave contributions (and any short-range correlations) are ignored in the energy eigenstructure. An alternate model for building the pseudo-eigenstructure based upon coupled harmonic oscillators that is more applicable at low temperatures can be found in reference \cite{yamada2018magnetization}.

\begin{figure}
\begin{center}
\includegraphics[scale=0.5]{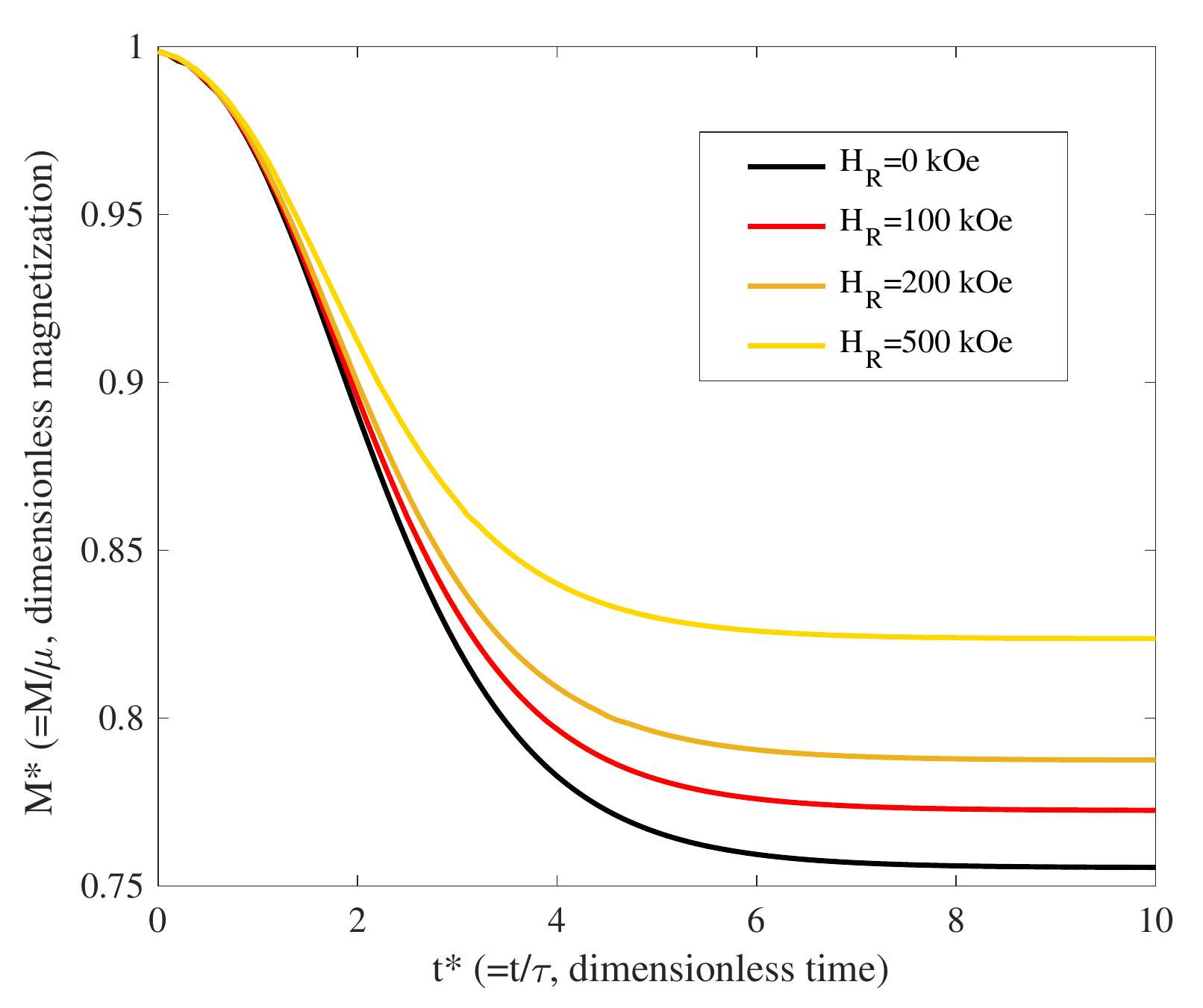}
\caption{\label{fig2:relaxation_magnetization} The calculated relaxation of magnetization in bcc-Fe at various external magnetic field strengths with $T_R=800$\;K using $N=10^6$. The initial states are prepared using $T_0=300$\;K and $H_0=0$\;kOe.  The magnetization, $M^*$, is a dimensionless magnetization normalized by the magnetic moment of iron, $M^*=M/\mu$, and $t^*$ is the dimensionless time normalized by the relaxation time, $t^*=t/\tau$. }
\end{center}
\end{figure}

The time-evolution process of magnetization can be calculated using the SEAQT equation of motion. Here, the relaxation process for a system interacting with a reservoir is investigated using Eq.\;(\ref{eq2:equation_motion_magnetization_heat}) where the initial probability distribution, $p^0_j$ is prepared using Eq.\;(\ref{eq2:canonical_distribution_magnetization}) by replacing $T^{\mbox{\footnotesize se}}$ and $H^{\mbox{\footnotesize se}}$ with $T_0$ and $H_0$. The calculated relaxation process at different external magnetic field strengths, $H_R=0$,\;100,\;200, and 500\;kOe, with $T_0=300$\;K, $H_0=0$\;kOe, and $T_R=800$\;K are shown in Fig.\;\ref{fig2:relaxation_magnetization}. Although the initial states are the same, the final states are different, each of which corresponds to the equilibrium values shown in Fig.\;\ref{fig2:stable_magnetization}, which are independently calculated from the canonical distribution, Eq.\;(\ref{eq2:canonical_distribution_magnetization}).

Note that as before the dimensionless time, $t^*$, normalized by the relaxation time, $\tau$, is used in the calculated relaxation processes. The relaxation time can be correlated with a real time by calibrating with either $ab$ $initio$ calculations \cite{beretta2014steepest,li2016generalized,li2018steepest,yamada2018method} or experimental data \cite{beretta2017steepest,li2018multiscale}. For the relaxation of magnetization, the experimental results of spin-pumping could be employed for real-time scaling \cite{yamada2018magnetization}.

\section{\label{chap2_sec:level6}Concluding comments}
In this paper, we have attempted to illustrate the methodology of applying the SEAQT framework to problems in materials science. With this framework, steepest entropy ascent dictates via an equation of motion the unique kinetic path a system follows from any initial non-equilibrium state to stable equilibrium. Since the method is based in Hilbert or Fock space with no explicit connection to a spatial or time scale, there are no inherent restrictions on the applicability of the SEAQT model in terms of system size or time.  For this reason, it is useful for multiscale calculations where a larger scale time-evolution process requires input from smaller scale behaviors within a single framework.

The SEAQT approach also has significant computational advantages relative to other computational methods. Many conventional computational tools in materials science require extensive information about the system being studied (e.g., the positions and momenta of particles and/or possible kinetic paths at each time-step), and this data is then updated in time through microscopic mechanics (e.g., molecular dynamics) or stochastic thermodynamics (e.g., kinetic Monte Carlo methods). Such methodologies place significant demands on computational resources such as the computational speed and data storage. The SEAQT framework is based upon a different paradigm. The kinetic path a system follows as its state evolves is found by simply solving $R$ first-order, ordinary differential equations (i.e., the SEAQT equation of motion) using energy and entropy as the fundamental state variables (where $R$ is the number of energy eigenlevels). For this reason, the computational cost in SEAQT modeling is remarkably small compared with conventional methods. For example, the kinetic paths shown in Fig.\;\ref{fig2:relaxation_magnetization} in Sec.\;\ref{chap2_sec:level5_2} ($R=555$ with $N=10^6$) were calculated in a few minutes on a laptop computer with $8$\;GB of memory.

As a final remark, there are three fronts where progress is needed to develop SEAQT applications for materials science. The first is a more elaborate description for the pseudo-eigenstructures.  Both the equilibrium and non-equilibrium properties calculated by the SEAQT method depend entirely on the accuracy of the pseudo-eigenstructure (or the underlying solid-state model). In the mean-field approach used in Sec.\;\ref{chap2_sec:level5_2}, for example, short-range correlations were ignored. For a more reliable description of material properties, short-range correlations could be added. This is especially relevant to alloy systems because it is known that short-range correlations between different atomic species can affect the kinetic paths of phase transformations. 

The second front is an extension of the method to heterogeneous systems. Although homogeneous systems have been assumed in references \cite{yamada2018method,yamada2018kineticpartI,yamada2018kineticpartII,yamada2018magnetization} (as well as in Sec.\;\ref{chap2_sec:level5_2} in this paper), most materials are highly heterogeneous at a mesoscopic scale. Lots of interesting behaviors are observed at this the scale (such as unique microstructures depending upon a stress field and lattice misfits). In order to describe the heterogeneous system, the construction of a network of local systems would be required as is done in reference \cite{li2018multiscale}. The third front is the coupling of different phenomena, which is something that is inherent to this framework. The topics investigated to date --- thermal expansion \cite{yamada2018method}, magnetization \cite{yamada2018magnetization}, and phase decomposition \cite{yamada2018kineticpartI,yamada2018kineticpartII}, for example, are not necessarily independent but may depend upon each other in nonlinear ways. For a complete description of solid-state phenomena, the inclusion of the coupling effects would be essential. To accomplish this aim within the SEAQT framework, a similar approach as that used to explore the coupled behavior between electrons and phonons \cite{li2018steepest} could be employed.

\section*{ACKNOWLEDGEMENT}
We acknowledge the National Science Foundation (NSF) for support through Grant DMR-1506936. \\

\begin{appendices}

\section*{\label{chap2_sec:level7}Appendix}

\section{\label{chap2_sec:level7_1}Quantum statistical mechanics and the quantum Boltzmann entropy }
Quantum statistical mechanics (QSM) is a bridge between quantum mechanics and thermodynamics as well as is SEAQT. Although both QSM and SEAQT are ensemble-based approaches, the concepts of ensemble are different in each framework. QSM uses a heterogeneous ensemble, whereas SEAQT is based on a homogeneous ensemble. Furthermore, while the SEAQT framework employs the von Neumann formula for the entropy, QSM employs the quantum Boltzmann entropy formula. In this appendix, the distinctions  between the ensembles and entropy formulas used are discussed. 

A homogeneous ensemble is an ensemble of identical systems that are identically prepared, while a heterogeneous ensemble is an ensemble of identical systems not identically prepared \cite{hatsopoulos1976-III,smith2012intrinsic}. In QSM, the state of a system is given as a weighted average of various states in a heterogeneous ensemble \footnote{Of course, as pointed out by Park \cite{park1968ensembles}, the use of a heterogeneous ensemble leads to the conclusion that knowledge of the state of the system, a bedrock of physical thought, is lost and all that can indeed be said is that the state of the ensemble and not that of the system is known. This is not the case for a homogeneous ensemble for which the state of the ensemble necessarily coincides with that of the system}. This causes a violation of the well-known second law of thermodynamics \cite{hatsopoulos1976-III,smith2012intrinsic} (i.e., no energy via a work interaction can be extracted from a system when the system is in a stable equilibrium state \cite{gyftopoulos2005thermodynamics}). In a heterogeneous ensemble, it is possible to extract work from the system in a stable equilibrium state because {\it some} of the states in the ensemble necessarily deviate from the average (stable equilibrium) --- a perpetual motion machine of the second kind \cite{gyftopoulos2005thermodynamics}. In the SEAQT framework, on the other hand, the state of a system is defined differently. It is given by an ensemble of energy eigenlevels for the system \cite{hatsopoulos1976-III,smith2012intrinsic} (rather than an ensemble of systems, each of which is in a different energy eigenlevel). Since the SEAQT framework does not average over a set of different states, it does not violate the second law of thermodynamics. 

Now, as to the von Neumann entropy formula, it satisfies all of the characteristics of the entropy required by thermodynamics \cite{gyftopoulos1997entropy,cubukcu1993thermodynamics}, while the quantum Boltzmann entropy formula makes entropy a statistical property (and not a fundamental one) that results from a loss of information. Nevertheless, QSM with the quantum Boltzmann entropy formula has produced great success in computational materials science. This suggests that there is a relationship between the two entropy formulae under some conditions. This relationship can be readily derived as follows. The von Neumann entropy is defined as 
\begin{equation}
s= - \sum_i p_i \mathrm{ln} \left( \frac{p_i}{g_i} \right)   \; , \label{eq2:neumann_entropy_formula}
\end{equation}
where $p_j$ and $g_j$ are, respectively, the occupation probability and the degeneracy in the $j^{th}$ energy eigenlevel, $\epsilon_j$ ($k_B$ is omitted here for simplicity). When the occupation probability, $p_j$, is localized at a single energy eigenlevel, $\epsilon_{j^*}$, the distribution is given as $p_{j^*}\approx1$ and $p_{j\neq j^*}\approx0$. Then, the von Neumann entropy formula, Eq.\;(\ref{eq2:neumann_entropy_formula}), becomes $s \approx \mathrm{ln} g_{j^*} $. This entropy corresponds with the quantum Boltzmann entropy formula, $s= \mathrm{ln} W$ (where $W$ is the number of complexions of the most probable state \cite{kittel1980thermal}), because both $g_{j^*}$ and $W$ represent the same physical quantity \footnote{The meaning of the term `complexions' corresponds to energy degeneracy used in the SEAQT framework} (even though they are based on different ensembles). Therefore, the Boltzmann entropy formula is valid when it is assumed that the occupation probability is highly localized at a given energy eigenlevel. Since, in QSM for a solid phase, it is assumed that the contribution of the most probable state is dominant compared with others when a stable equilibrium state is reached, the use of the quantum Boltzmann entropy formula may be justified. However, the assumption is rigorously exact only for an infinite, bulk sample \cite{girifalco2003statistical}.

\section{\label{chap2_sec:level7_2}The concept of hypo-equilibrium states }
The concept of hypo-equilibrium states developed by Li and von Spakovsky \cite{li2016steepest,li2016generalized} within the SEAQT theoretical framework provides simple relaxation patterns for systems. The concept makes the SEAQT equation of motion quite simple and tractable. Here, the basic idea is described and non-equilibrium intensive properties are defined.  

Using the steepest-entropy-ascent principle, it has been proven that when an initial state is divided into $M$ subspaces, each of which is in a canonical distribution (this is called a $M^{th}$-order hypo-equilibirum state), the system remains in a $M^{th}$-order hypo-equilibirum states during the entire time-evolution process \cite{li2016steepest}. Therefore, the probability distribution in the each subspace can be described as
\begin{equation}
\begin{split}
p^K_j (t^*) &= p^K(t^*) \frac{g^K_j \; \mathrm{exp}[ -\beta (t^*) \epsilon^K_j ] }{\sum\limits_i g^K_i \;\mathrm{exp}[ -\beta(t^*) \epsilon^K_i ]} \\
&= p^K(t^*) \; \frac{g^K_j \; \mathrm{exp}[ -\beta^K (t^*) \epsilon^K_j ] }{ Z^K(t^*)}  \; , \label{eq2:canonical_distribution_time_dependent_multiple_systems}
\end{split}
\end{equation}
where $\beta (t^*)=1/k_BT(t^*)$, $p^K_j$ and $g^K_j$ are, respectively, the occupation probability and the degeneracy in the energy eigenlevel $\epsilon^K_j$ in the $K^{th}$ subspace and $p^K$ is the mole fraction of the subspace. Any state can be represented using the canonical distribution by properly dividing the system into subspaces. The canonical distribution in a non-equilibrium state allows us to define intensive properties (e.g., temperature) in the non-equilibrium region. Intensive properties defined this way are fundamental \cite{li2016steepest} unlike a phenomenological temperature defined, for example, via the kinetic energy of the particles, $E=\frac{3}{2}k_BT$. \cite{casas1994nonequilibrium} The definitions and uses of the non-equilibrium intensive properties are found in references \cite{li2017study,li2018steepest,yamada2018magnetization}.

This is also true for subsystems that constitute a composite system \cite{li2016steepest}. In Fig.\;\ref{fig2:isolated_systems}\;(b), for example, two systems interacting via a heat interaction are considered with no mass exchange, i.e., $p^A(t^*)=p^B(t^*)=1$. Therefore, if the initial states of the each (sub) system, $A$ and $B$, are described by a canonical distribution, the time-evolution of occupation probabilities in the each subsystem are given as  
\begin{equation}
p^{A(B)}_j (t^*)  = \frac{g^{A(B)}_j \; \mathrm{exp}[ -\beta (t^*) \epsilon^{A(B)}_j ] }{ Z^{A(B)}(t^*)}  \; .  \label{eq2:canonical_distribution_time_dependent}
\end{equation}

Furthermore, using the concept of hypo-equilibrium states, the substitution of $C_3/C_1 \equiv \beta$ in Eq.\;(\ref{eq2:equation_of_motion_interacting_simplified}) can be justified. With the use of Eq.\;(\ref{eq2:canonical_distribution_time_dependent}), Eq.\;(\ref{eq2:equation_of_motion_interacting_simplified}) is written as
\begin{equation}
\begin{split}
\frac{dp^A_j}{dt^*} &=  p^A_j \left[ (s^A_j -  \left< s \right>^A)  - (\epsilon^A_j - \left< e \right>^A) \beta  \right]  \\
 \Rightarrow \;\; \frac{d (\mathrm{ln}p^A_j)}{dt^*}  &=  \left[ (s^A_j -  \left< s \right>^A) - (\epsilon^A_j - \left< e \right>^A) \beta  \right]  \\
  \Rightarrow\;\;  \frac{d}{dt^*} & \left( - \beta^A(t^*)\epsilon^A_j  -\mathrm{ln}Z^A(t^*) \right) \\
 & \qquad \qquad =  (\epsilon^A_j -  \left< e \right>^A )  (\beta^A(t^*) - \beta)  \; , 
\end{split}
\label{eq2:equation_of_motion_interacting_simplified_with_hypoequi}
\end{equation}
where the following relations are used:
\begin{equation}
\begin{split}
 \mathrm{ln} \left( \frac{d p^A_j }{dt^*} \right) & = \mathrm{ln} \left( \frac{d (\mathrm{ln} p^A_j) }{dt^*}  \frac{d p^A_j }{d (\mathrm{ln} p^A_j)}  \right)  \\
& = \mathrm{ln} \left( \frac{d (\mathrm{ln} p^A_j) }{dt^*}  \right) + \mathrm{ln} \left( \frac{d p^A_j }{d (\mathrm{ln} p^A_j)}  \right)  \\
& = \mathrm{ln} \left(   \frac{d(\mathrm{ln} p^A_j) }{dt^*}  \right) + \mathrm{ln} p^A_j  \\
 \Rightarrow \;\; \frac{d(\mathrm{ln} p^A_j) }{dt^*} &  = \mathrm{exp} \left[   \mathrm{ln} \left( \frac{d p^A_j }{dt^*} \right)  - \mathrm{ln} p^A_j   \right] =  \frac{1}{p^A_j} \frac{d p^A_j }{dt^*}
 \; , 
\end{split}
\label{eq2:log_relation} 
\end{equation}
and
\begin{equation}
\begin{split}
s^A_j &= -\mathrm{ln}\frac{p^A_j}{g^A_j} = \beta^A(t^*) \epsilon^A_j + \mathrm{ln}Z^A(t^*) \\
\left< s \right>^A &= - \sum_i p^A_i \mathrm{ln} \frac{p^A_i}{g^A_i} = \beta^A(t^*) \left< e \right>^A + \mathrm{ln} Z^A(t^*) 
 \; .
\end{split}
\label{eq2:equation_of_motion_interacting_simplified_with_hypoequi_s}
\end{equation}
Subtracting Eq.\;(\ref{eq2:equation_of_motion_interacting_simplified_with_hypoequi}) for the $i^{th}$ and $j^{th}$ energy eigenlevels yields \cite{li2016steepest}
\begin{equation}
\begin{split}
 \frac{d}{dt^*} \left[ - \beta^A(t^*) (\epsilon^A_i - \epsilon^A_j ) \right]  & = (\epsilon^A_i -  \epsilon^A_j ) (\beta^A(t^*) - \beta) \\
 \Rightarrow \;   \frac{d\beta^A(t^*)}{dt^*} & = - (\beta^A(t^*) - \beta)
 \; . 
\end{split}
\label{eq2:equation_of_motion_interacting_simplified_with_hypoequi_subtract}
\end{equation}
This is the equation of motion for the intensive property, $\beta^A$. At stable equilibrium, $ d\beta^A(t^*)/dt^* \rightarrow 0$, which corresponds to the condition, $\beta^A(t^*) = \beta$. Therefore, $\beta\;(\equiv C_3/C_1)$ is considered to be $1/k_BT$ as defined in Eq.\;(\ref{eq2:canonical_distribution_time_dependent_multiple_systems}). When system\;$B$ is viewed as a heat reservoir, $\beta$ is replaced by $\beta^R$ and Eq.\;(\ref{eq2:equation_of_motion_interacting_simplified_with_hypoequi_subtract}) becomes
\begin{equation}
\begin{split}
\frac{d\beta(t^*)}{dt^*} = - (\beta(t^*) - \beta^R)  \; , 
\end{split}
\label{eq2:equation_of_motion_temperature_reservoir}
\end{equation}
where the superscripts, $A$, are removed. Therefore, the time-evolution of a system that interacts with a heat reservoir can be determined readily from Eqs.\;(\ref{eq2:canonical_distribution_time_dependent}) and (\ref{eq2:equation_of_motion_temperature_reservoir}) if the initial state of the system is described by a canonical distribution. A more detailed discussion about hypo-equilibrium states and a more general case (e.g., for heat and mass diffusion between interacting systems) can be found in reference \cite{li2016generalized}.

\section{\label{chap2_sec:level7_3}Spin energy using the Ising model with the mean-field approximation}
An approximate energy in a spin system is derived using the Ising model with a mean-field approximation in this appendix. The energy in a spin system is given by Eq.\;(\ref{eq2:energy_pair_interaction}) by taking into account only the first-nearest-neighbor pair interactions. Using the mean-field approximation, which does not include any short-range correlations between spins, the pair probabilities in Eq.\;(\ref{eq2:energy_pair_interaction}) are given by a product of probabilities of up- and/or down-spins as
\[
\begin{array}{c c}
y_{\uparrow \uparrow}=x_\uparrow x_\uparrow  \;,  \;\;\; 
&
y_{\uparrow \downarrow}=x_\uparrow x_\downarrow  \; , \\ \\
y_{\downarrow \uparrow}=x_\downarrow x_\uparrow \; , \;\;\; 
&
y_{\downarrow \downarrow}=x_\downarrow x_\downarrow  \; , \\ \\
\end{array}
\]
where $x_\uparrow$ and $x_\downarrow$ are, respectively, the probability of up-spins and down-spins in a system. Then, Eq.\;(\ref{eq2:energy_pair_interaction}) can be expanded as 
\begin{equation}
\begin{split}
E&= \frac{1}{2} Nz \left( e_{\uparrow \uparrow} x_\uparrow x_\uparrow + 2 e_{\uparrow \downarrow} x_\uparrow x_\downarrow + e_{\downarrow \downarrow} x_\downarrow x_\downarrow  \right)    \\
&= \frac{1}{2} Nz \left[ e_{\uparrow \uparrow} (1-c)^2 + 2 e_{\uparrow \downarrow} c(1-c) + e_{\downarrow \downarrow} c^2  \right]   
\; ,   
\end{split}
\label{eq2:energy_pair_interaction_mean_field_0}
\end{equation}
where $x_\uparrow$ and $x_\downarrow$ are replaced as $x_\uparrow=1-c$ and $x_\downarrow=c$ by defining the fraction of down-spins, $c$. Now, the reference energy of Eq.\;(\ref{eq2:energy_pair_interaction_mean_field_0}) is set to the line connecting two energies of all up-spins ($c=0$) or all down-spins ($c=1$) as 
\begin{equation}
\Delta E(c) = E(c) - \frac{E(1.0)-E(0.0)}{1.0-0.0} c \\
\; .     \label{eq2:change_energy_reference}
\end{equation}
Thus, the energy becomes
\begin{equation}
\begin{split}
\Delta E(c) & = \frac{1}{2} Nz  (2e_{\uparrow \downarrow}- e_{\uparrow \uparrow} - e_{\downarrow \downarrow}) c (1-c)   \\
& = \frac{1}{2} Nz \; J_{\mbox{\scriptsize eff}}\; c(1-c)  
\; ,     
\end{split}
\label{eq2:energy_pair_interaction_mean_field_change_reference}
\end{equation}
where $J_{\mbox{\scriptsize eff}} $ is the effective interaction energy defined as $J_{\mbox{\scriptsize eff}} \equiv 2e_{\uparrow \downarrow}- e_{\uparrow \uparrow} - e_{\downarrow \downarrow}$. 

The effective interaction energy, $J_{\mbox{\scriptsize eff}} $, can be determined either from $ab$ $initio$ calculations \cite{pajda2001ab} or from experiments. Here, it is roughly estimated using the experimentally measured Curie temperature of iron, $T_c=1043$\;K. \cite{kittel1966introduction} The Helmholtz free energy of the spin system is given by
\begin{equation}
\begin{split}
F& = E-TS =E-k_BT \mathrm{ln}W  \\
& = \frac{1}{2} Nz \; J_{\mbox{\scriptsize eff}}\; c(1-c)  - k_B T \mathrm{ln} \frac{N !}{(N(1-c))! (Nc)!} 
\; ,   
\end{split}
\label{eq2:free_energy_spin_system}
\end{equation}
where Eq.\;(\ref{eq2:energy_pair_interaction_mean_field_change_reference}) is used in the energy term and the quantum Boltzmann entropy formula is employed. Applying Stirling's formula, $\mathrm{ln}x ! \approx x\mathrm{ln}x -x$, Eq.\;(\ref{eq2:free_energy_spin_system}) becomes
\begin{equation}
\begin{split}
F  = \frac{Nz}{2} \; & J_{\mbox{\scriptsize eff}} \; c(1-c)  \\
& - Nk_B T  \left[ c\mathrm{ln}(1-c) -c \mathrm{ln} c - \mathrm{ln}(1-c)  \right]
\; .     
\end{split}
\label{eq2:free_energy_spin_system_stirling}
\end{equation}
It is expected that the second derivative of the free energy in terms of the fraction of down-spins, $c$, becomes zero at the Curie temperature, $T_c$, and $c=0.5$, i.e., $(d^2F/dc^2)_{c=0.5}=0$. Using this relation, $J_{\mbox{\scriptsize eff}} $ is derived as
\begin{equation}
\begin{split}
T_c = \frac{z c(1-c) J_{\mbox{\scriptsize eff}} }{ k_B} =  \frac{z J_{\mbox{\scriptsize eff}} }{ 4 k_B} \Rightarrow \;  J_{\mbox{\scriptsize eff}} = \frac{4 k_B T_c}{z}
\; .  
\end{split}
\label{eq2:effective_interaction_energy}
\end{equation}
Since $T_c=1043$\;K \cite{kittel1966introduction} and $z=8$ for bcc-Fe, the effective interaction energy becomes $J_{\mbox{\scriptsize eff}}=7.2 \times 10 ^{-22}$ (J/atom).

Note that an $ad$ $hoc$ assumption is used here for the estimation of $J_{\mbox{\scriptsize eff}}$, i.e., the second derivative of $F$ in terms of $c$ becomes zero at $T=T_c$ and $c=0.5$. A more reliable approach for estimating $J_{\mbox{\scriptsize eff}}$ can be found in references \cite{girifalco2003statistical,kittel1980thermal,aharoni2000introduction}. Furthermore, the free-energy analysis, Eq.\;(\ref{eq2:free_energy_spin_system}), is used here just for the estimation of $J_{\mbox{\scriptsize eff}}$. In the SEAQT framework, no free-energy functions are used because these functions are strictly applicable only at stable equilibrium.

\section{\label{chap2_sec:level7_4}Computational Tips}
There are many computational tools that can be used for SEAQT modeling. The calculations shown in Sec.\;\ref{chap2_sec:level5_2} are conducted using Mathematica\;(11.2.0.0) and MATLAB\;(R2017a). They are used for the calculations of the energy eigenstructure and to solve the equation of motion, respectively.

The relaxation processes are numerically calculated with an ordinary differential equation (ODE) solver in MATLAB (e.g., ode45 and ode15). In MATLAB, however, very small/large values are treated as zero/infinity, and the ODEs (or the SEAQT equation of motion) cannot be solved. This becomes a problem when the number of energy eigenlevels and/or the degeneracy of the energy eigenlevels become very large (because the occupation probability, $p_j$, and the degeneracy, $g_j$, become $p_j \rightarrow 0$ and $g_j \rightarrow$\;infinity, respectively). The way to avoid the issue is described in this appendix.

The most straightforward approach to circumvent the problem is to use the logarithm of $p_j$ and $g_j$. For example, the SEAQT equation of motion, Eq.\;(\ref{eq2:equation_motion_magnetization_heat}), can be rewritten using the logarithm as
\begin{equation}
 \frac{d(\mathrm{ln} p_j) }{dt^*} =  \left[ \left( s_j - \langle s \rangle \right)  - \left( \epsilon_j - \left< e \right> \right) \beta^R + \left( m_j- \langle m \rangle \right) \gamma^R  \right] \; ,    \label{eq2:equation_motion_magnetization_heat_log}
\end{equation}
where the relation, Eq.\;(\ref{eq2:log_relation}), is used. Note that all $p_j$ and $g_j$ in Eq.\;(\ref{eq2:equation_motion_magnetization_heat_log}) also need to be converted to the logarithmic forms in the code such that
\begin{equation}
\begin{split}
\;\; \;\; s_j &=-  \mathrm{ln} \frac{p_j}{g_j}  =  \mathrm{ln} g_j - \mathrm{ln} p_j   \\
\;\;  \langle s \rangle & =- \sum\limits_{i} p_i \mathrm{ln} \frac{p_i}{g_i}  =\sum_i \mathrm{exp} \left[ \mathrm{ln}p_i + \mathrm{ln} (\mathrm{ln} g_i - \mathrm{ln} p_i)  \right]  \\
\;\;  \langle e \rangle & = \sum\limits_{i} \epsilon_i p_i  =\sum_i \mathrm{exp} \left( \mathrm{ln}\epsilon_i + \mathrm{ln} p_i  \right)  \\
\;\;  \langle m \rangle & = \sum\limits_{i} m_i p_i  =\sum_i \mathrm{exp} \left( \mathrm{ln} m_i + \mathrm{ln} p_i  \right)  
\; .   
\end{split}
\label{eq2:log_other_property} 
\end{equation}

A similar computational issue is faced with calculating stable equilibrium states for a system that has a huge number of energy eigenlevels or an enormous degeneracy. A stable equilibrium state is determined from a canonical distribution, e.g., Eqs.\;(\ref{eq2:canonical_distribution}) and (\ref{eq2:canonical_distribution_magnetization}). In the canonical distribution, the problem is evaluating the partition function, e.g., $Z \equiv \sum_i g_i \mathrm{exp}( -\epsilon_i / k_B T ) $, because some terms in the partition function are converted to infinity by some software. The problem can be avoided using the logarithms as well. The partition function can be expanded as   
\begin{equation}
\begin{split}
&Z \equiv \sum_i g_i e^{ - \beta \epsilon_i } = X_1 + X_2 + X_3 + ... +X_R  \\
&= X_{\mathrm{max}} \left( \frac{X_1}{X_{\mathrm{max}}} + \frac{X_2}{X_{\mathrm{max}}}+ ... + 1 + ....+ \frac{X_R}{X_{\mathrm{max}}} \right)   
\; ,  
\end{split}
\label{eq2:partition_function_expand} 
\end{equation}
where $X_j \equiv g_j  e^{ - \beta \epsilon_j } $, $R$ is the number of energy eigenlevels, and $X_{\mathrm{max}}$ is the maximum $X_j$ in the expansion. Using the logarithm of $X_j$, i.e., $\mathrm{ln}X_j = \mathrm{ln} g_j - \beta \epsilon_j $, Eq.\;(\ref{eq2:partition_function_expand}) is written as
\begin{equation}
\begin{split}
\mathrm{ln} & Z  = \mathrm{ln} X_{\mathrm{max}} + \mathrm{ln} \left( \frac{X_1}{X_{\mathrm{max}}} + \frac{X_2}{X_{\mathrm{max}}}+ ... + \frac{X_R}{X_{\mathrm{max}}} \right)  \\
= & \mathrm{ln} X_{\mathrm{max}}  + \mathrm{ln} [ \mathrm{exp}(\mathrm{ln} X_1  -  \mathrm{ln} X_{\mathrm{max}})  \\
\;\;\; & +  \mathrm{exp}(\mathrm{ln} X_2  -  \mathrm{ln} X_{\mathrm{max}}) + ... +  \mathrm{exp}(\mathrm{ln} X_R  -  \mathrm{ln} X_{\mathrm{max}})  ]
\; .  
\end{split}
\label{eq2:partition_function_log} 
\end{equation}
The canonical distribution, $p_j=g_j  e^{ - \beta \epsilon_j }/Z$, can then be calculated using the logarithms as
\begin{equation}
\begin{split}
&\mathrm{ln} p_j  = \mathrm{ln} g_j - \beta \epsilon_j - \mathrm{ln} Z  \\
 \Rightarrow \;\;\; & \ p_j  = \mathrm{exp} \left( \mathrm{ln} g_j - \beta \epsilon_j - \mathrm{ln} Z \right)
\; . 
\end{split}
\label{eq2:canonical_log} 
\end{equation}

Although the degeneracy, $g_j$, in Sec.\;\ref{chap2_sec:level5_2} are directly evaluated from Eq.\;(\ref{eq2:degeneracy_pseud_spin_flip}) using Mathematica, they can be  estimated simply using the following relation:
\begin{equation}
\begin{split}
& \;\;\;\quad\quad\quad g_j =\frac{N !}{(N(1-c_j))!  (Nc_j)!} \;     \\
%\Rightarrow 
%& \;\;\; 
&\mathrm{ln} g_j \approx N \cdot \mathrm{ln} \left[ c_j  \mathrm{ln}(1-c_j) - c_j  \mathrm{ln} c_j - \mathrm{ln} (1-c_j) \right]
\end{split}
\label{eq2:degeneracy_ising_stirling}
\end{equation}
where the Stirling formula, $\mathrm{ln}\,x ! \approx x\mathrm{ln}\,x -x$, is employed. From this relation, it is evident that $\mathrm{ln}\,g_j$ is simply proportional to the number of particles, $N$. Therefore, once the degeneracies for a system composed of a small number of particles (say, $N=N^S$), i.e., $\mathrm{ln}\,g_j^S$, are calculated from Eq.\;(\ref{eq2:degeneracy_pseud_spin_flip}), the degeneracies for a large number of particles (say, $N=N^L$), i.e., $\mathrm{ln}\,g_j^L$, can be determined from $\mathrm{ln}\,g_j^L=\frac{N^L}{N^S} \, \mathrm{ln}\,g_j^S$.

\end{appendices}

\bibliographystyle{ieeetr}
\bibliography{ref}

\begin{thebibliography}{10}

\bibitem{maddox1985uniting}
J.~Maddox, ``Uniting mechanics and statistics,'' {\em Nature}, vol.~316, p.~11,
  1985.

\bibitem{hatsopoulos1976-I}
G.~N. Hatsopoulos and E.~P. Gyftopoulos, ``A unified quantum theory of
  mechanics and thermodynamics. {P}art {I}. postulates,'' {\em Foundations of
  Physics}, vol.~6, no.~1, pp.~15--31, 1976.

\bibitem{hatsopoulos1976-IIa}
G.~N. Hatsopoulos and E.~P. Gyftopoulos, ``A unified quantum theory of
  mechanics and thermodynamics. {P}art {II}a. {A}vailable energy,'' {\em
  Foundations of Physics}, vol.~6, no.~2, pp.~127--141, 1976.

\bibitem{hatsopoulos1976-IIb}
G.~N. Hatsopoulos and E.~P. Gyftopoulos, ``A unified quantum theory of
  mechanics and thermodynamics. {P}art {II}b. {S}table equilibrium states,''
  {\em Foundations of Physics}, vol.~6, no.~4, pp.~439--455, 1976.

\bibitem{hatsopoulos1976-III}
G.~N. Hatsopoulos and E.~P. Gyftopoulos, ``A unified quantum theory of
  mechanics and thermodynamics. {P}art {III}. {I}rreducible quantal
  dispersions,'' {\em Foundations of Physics}, vol.~6, no.~5, pp.~561--570,
  1976.

\bibitem{beretta2005generalPhD}
G.~P. Beretta, {\em On the general equation of motion of quantum thermodynamics
  and the distinction between quantal and nonquantal uncertainties}.
\newblock PhD thesis, Massachusetts Institute of Technology, 1981.

\bibitem{beretta1984quantum}
G.~P. Beretta, E.~P. Gyftopoulos, J.~L. Park, and G.~N. Hatsopoulos, ``Quantum
  thermodynamics. {A} new equation of motion for a single constituent of
  matter,'' {\em Il Nuovo Cimento B}, vol.~82, no.~2, pp.~169--191, 1984.

\bibitem{beretta1985quantum}
G.~P. Beretta, E.~P. Gyftopoulos, and J.~L. Park, ``Quantum thermodynamics. {A}
  new equation of motion for a general quantum system,'' {\em Il Nuovo Cimento
  B}, vol.~87, no.~1, pp.~77--97, 1985.

\bibitem{beretta2006nonlinear}
G.~P. Beretta, ``{N}onlinear model dynamics for closed-system, constrained,
  maximal-entropy-generation relaxation by energy redistribution,'' {\em
  Physical Review E}, vol.~73, no.~2, p.~026113, 2006.

\bibitem{beretta2009nonlinear}
G.~P. Beretta, ``{N}onlinear quantum evolution equations to model irreversible
  adiabatic relaxation with maximal entropy production and other nonunitary
  processes,'' {\em Reports on Mathematical Physics}, vol.~64, no.~1/2,
  pp.~139--168, 2009.

\bibitem{beretta2014steepest}
G.~P. Beretta, ``{S}teepest entropy ascent model for far-nonequilibrium
  thermodynamics: {U}nified implementation of the maximum entropy production
  principle,'' {\em Physical Review E}, vol.~90, no.~4, p.~042113, 2014.

\bibitem{von2014some}
M.~R. von Spakovsky and J.~Gemmer, ``Some trends in quantum thermodynamics,''
  {\em Entropy}, vol.~16, no.~6, pp.~3434--3470, 2014.

\bibitem{montefusco2015essential}
A.~Montefusco, F.~Consonni, and G.~P. Beretta, ``{E}ssential equivalence of the
  general equation for the nonequilibrium reversible-irreversible coupling
  ({GENERIC}) and steepest-entropy-ascent models of dissipation for
  nonequilibrium thermodynamics,'' {\em Physical Review E}, vol.~91, no.~4,
  p.~042138, 2015.

\bibitem{cano2015steepest}
S.~Cano-Andrade, G.~P. Beretta, and M.~R. von Spakovsky,
  ``{S}teepest-entropy-ascent quantum thermodynamic modeling of decoherence in
  two different microscopic composite systems,'' {\em Physical Review A},
  vol.~91, no.~1, p.~013848, 2015.

\bibitem{smith2016comparing}
C.~E. Smith, ``Comparing the {M}odels of {S}teepest {E}ntropy {A}scent
  {Q}uantum {T}hermodynamics, {M}aster {E}quation and the {D}ifference
  {E}quation for a {S}imple {Q}uantum {S}ystem {I}nteracting with
  {R}eservoirs,'' {\em Entropy}, vol.~18, no.~5, p.~176, 2016.

\bibitem{beretta2017steepest}
G.~P. Beretta, O.~Al-Abbasi, and M.~R. von Spakovsky, ``Steepest-entropy-ascent
  nonequilibrium quantum thermodynamic framework to model chemical reaction
  rates at an atomistic level,'' {\em Physical Review E}, vol.~95, no.~4,
  p.~042139, 2017.

\bibitem{li2016steepest}
G.~Li and M.~R. von Spakovsky, ``{S}teepest-entropy-ascent quantum
  thermodynamic modeling of the relaxation process of isolated chemically
  reactive systems using density of states and the concept of hypoequilibrium
  state,'' {\em Physical Review E}, vol.~93, no.~1, p.~012137, 2016.

\bibitem{li2016generalized}
G.~Li and M.~R. von Spakovsky, ``Generalized thermodynamic relations for a
  system experiencing heat and mass diffusion in the far-from-equilibrium realm
  based on steepest entropy ascent,'' {\em Physical Review E}, vol.~94, no.~3,
  p.~032117, 2016.

\bibitem{li2016modeling}
G.~Li and M.~R. von Spakovsky, ``Modeling the nonequilibrium effects in a
  nonquasi-equilibrium thermodynamic cycle based on steepest entropy ascent and
  an isothermal-isobaric ensemble,'' {\em Energy}, vol.~115, pp.~498--512,
  2016.

\bibitem{li2016steepest2}
G.~Li and M.~R. von Spakovsky, ``Steepest-entropy-ascent quantum thermodynamic
  modeling of the far-from-equilibrium interactions between nonequilibrium
  systems of indistinguishable particle ensembles,'' {\em arXiv preprint
  arXiv:1601.02703}, 2016.

\bibitem{li2017study}
G.~Li and M.~R. von Spakovsky, ``Study of {N}onequilibrium {S}ize and
  {C}oncentration {E}ffects on the {H}eat and {M}ass {D}iffusion of
  {I}ndistinguishable {P}articles using {S}teepest-{E}ntropy-{A}scent {Q}uantum
  {T}hermodynamics,'' {\em Journal of Heat Transfer}, vol.~139, no.~12,
  p.~122003, 2017.

\bibitem{li2018multiscale}
G.~Li, M.~R. von Spakovsky, F.~Shen, and K.~Lu, ``Multiscale {T}ransient and
  {S}teady-{S}tate {S}tudy of the {I}nfluence of {M}icrostructure {D}egradation
  and {C}hromium {O}xide {P}oisoning on {S}olid {O}xide {F}uel {C}ell {C}athode
  {P}erformance,'' {\em Journal of Non-Equilibrium Thermodynamics}, vol.~43,
  no.~1, pp.~21--42, 2018.

\bibitem{li2018steepest}
G.~Li, M.~R. von Spakovsky, and C.~Hin, ``Steepest entropy ascent quantum
  thermodynamic model of electron and phonon transport,'' {\em Physical Review
  B}, vol.~97, no.~2, p.~024308, 2018.

\bibitem{yamada2018method}
R.~Yamada, M.~R. von Spakovsky, and W.~T. Reynolds~Jr., ``A method for
  predicting non-equilibrium thermal expansion using steepest-entropy-ascent
  quantum thermodynamics,'' {\em Journal of Physics: Condensed Matter},
  vol.~30, no.~32, p.~325901, 2018.

\bibitem{yamada2018kineticpartI}
R.~Yamada, M.~R. von Spakovsky, and W.~T. Reynolds~Jr., ``Kinetic {P}athways of
  {P}hase {D}ecomposition {U}sing {S}teepest-{E}ntropy-{A}scent {Q}uantum
  {T}hermodynamics {M}odeling. {P}art\;{I}: {C}ontinuous and {D}iscontinuous
  {T}ransformations,'' {\em arXiv preprint arXiv:1809.10627}, 2018.

\bibitem{yamada2018kineticpartII}
R.~Yamada, M.~R. von Spakovsky, and W.~T. Reynolds~Jr., ``{K}inetic {P}athways
  of {P}hase {D}ecomposition {U}sing {S}teepest-{E}ntropy-{A}scent {Q}uantum
  {T}hermodynamics {M}odeling. {P}art\;{II}: {P}hase {S}eparation and
  {O}rdering,'' {\em arXiv preprint arXiv:1809.10633}, 2018.

\bibitem{yamada2018magnetization}
R.~Yamada, M.~R. von Spakovsky, and W.~T. Reynolds~Jr., ``Low-temperature
  {A}tomistic {S}pin {R}elaxations and {N}on-equilibrium {I}ntensive
  {P}roperties {U}sing {S}teepest-{E}ntropy-{A}scent {Q}uantum {T}hermodynamics
  {M}odeling,'' {\em arXiv preprint arXiv:1809.10619}, 2018.

\bibitem{kittel1980thermal}
C.~Kittel and H.~Kroemer, {\em Thermal physics}.
\newblock W. H. Freeman, 2nd~ed., 1980.

\bibitem{girifalco2003statistical}
L.~A. Girifalco, {\em Statistical mechanics of solids}, vol.~58.
\newblock OUP USA, 2003.

\bibitem{smith2012intrinsic}
C.~E. Smith, {\em {I}ntrinsic {Q}uantum {T}hermodynamics: {A}pplication to
  hydrogen storage on a carbon nanotube and theoretical consideration of
  non-work interactions}.
\newblock PhD thesis, Virginia Polytechnic Institute and State University,
  2012.

\bibitem{doebner1992general}
H.-D. Doebner and G.~A. Goldin, ``On a general nonlinear {S}chr{\"o}dinger
  equation admitting diffusion currents,'' {\em Physics Letters A}, vol.~162,
  no.~5, pp.~397--401, 1992.

\bibitem{schuch2010pythagorean}
D.~Schuch, ``Pythagorean quantization, action(s) and the arrow of time,'' in
  {\em Journal of Physics: Conference Series}, vol.~237, p.~012020, IOP
  Publishing, 2010.

\bibitem{gemmer2004quantum}
J.~Gemmer, M.~Michel, and G.~Mahler, ``Quantum {T}hermodynamics: {E}mergence of
  {T}hermodynamic {B}ehavior {W}ithin {C}omposite {Q}uantum {S}ystems, volume
  657 of {L}ecture {N}otes in {P}hysics,'' 2004.

\bibitem{gemmer2009quantum}
J.~Gemmer, M.~Michel, and G.~Mahler, ``Quantum {T}hermodynamics: {E}mergence of
  {T}hermodynamic {B}ehavior {W}ithin {C}omposite {Q}uantum {S}ystems, volume
  784 of {L}ecture {N}otes in {P}hysics,'' 2009.

\bibitem{zurek1994decoherence}
W.~H. Zurek and J.~P. Paz, ``Decoherence, chaos, and the second law,'' {\em
  Physical Review Letters}, vol.~72, no.~16, p.~2508, 1994.

\bibitem{bathe2007finite}
K.-J. Bathe, ``Finite element method,'' {\em Wiley encyclopedia of computer
  science and engineering}, pp.~1--12, 2007.

\bibitem{dhatt2012finite}
G.~Dhatt, E.~Lefran{\~A}, and G.~Touzot, {\em Finite element method}.
\newblock John Wiley \& Sons, 2012.

\bibitem{chen2002phase}
L.-Q. Chen, ``Phase-field models for microstructure evolution,'' {\em Annual
  review of materials research}, vol.~32, no.~1, pp.~113--140, 2002.

\bibitem{moelans2008introduction}
N.~Moelans, B.~Blanpain, and P.~Wollants, ``An introduction to phase-field
  modeling of microstructure evolution,'' {\em Calphad}, vol.~32, no.~2,
  pp.~268--294, 2008.

\bibitem{binder2004molecular}
K.~Binder, J.~Horbach, W.~Kob, W.~Paul, and F.~Varnik, ``Molecular dynamics
  simulations,'' {\em Journal of Physics: Condensed Matter}, vol.~16, no.~5,
  p.~S429, 2004.

\bibitem{voter2007introduction}
A.~F. Voter, ``Introduction to the kinetic {M}onte {C}arlo method,'' in {\em
  Radiation effects in solids}, pp.~1--23, Springer, 2007.

\bibitem{lesar2013introduction}
R.~LeSar, {\em Introduction to computational materials science: fundamentals to
  applications}.
\newblock Cambridge University Press, 2013.

\bibitem{onodera2014recent}
H.~Onodera, T.~Abe, M.~Shimono, and T.~Koyama, ``Recent {A}dvances in
  {C}omputational {M}aterials {S}cience,'' {\em TETSU TO HAGANE-JOURNAL OF THE
  IRON AND STEEL INSTITUTE OF JAPAN}, vol.~100, no.~10, pp.~1207--1219, 2014.

\bibitem{weinan2011principles}
E.~Weinan, {\em Principles of multiscale modeling}.
\newblock Cambridge University Press, 2011.

\bibitem{hoyt2002atomistic}
J.~J. Hoyt and M.~Asta, ``Atomistic computation of liquid diffusivity,
  solid-liquid interfacial free energy, and kinetic coefficient in {A}u and
  {A}g,'' {\em Physical Review B}, vol.~65, no.~21, p.~214106, 2002.

\bibitem{vaithyanathan2004multiscale}
V.~Vaithyanathan, C.~Wolverton, and L.-Q. Chen, ``Multiscale modeling of
  $\theta$' precipitation in {A}l--{C}u binary alloys,'' {\em Acta Materialia},
  vol.~52, no.~10, pp.~2973--2987, 2004.

\bibitem{yamanaka2008coupled}
A.~Yamanaka, T.~Takaki, and Y.~Tomita, ``Coupled simulation of microstructural
  formation and deformation behavior of ferrite--pearlite steel by phase-field
  method and homogenization method,'' {\em Materials Science and Engineering:
  A}, vol.~480, no.~1-2, pp.~244--252, 2008.

\bibitem{fromm2012linking}
B.~S. Fromm, K.~Chang, D.~L. McDowell, L.-Q. Chen, and H.~Garmestani, ``Linking
  phase-field and finite-element modeling for process--structure--property
  relations of a {N}i-base superalloy,'' {\em Acta Materialia}, vol.~60,
  no.~17, pp.~5984--5999, 2012.

\bibitem{balluffi2005kinetics}
R.~W. Balluffi, S.~M. Allen, and W.~C. Carter, {\em Kinetics of materials}.
\newblock John Wiley \& Sons, 2005.

\bibitem{gyftopoulos1997entropy}
E.~P. Gyftopoulos and E.~Cubukcu, ``Entropy: thermodynamic definition and
  quantum expression,'' {\em Physical Review E}, vol.~55, no.~4, p.~3851, 1997.

\bibitem{cubukcu1993thermodynamics}
E.~Cubukcu, {\em {T}hermodynamics as a non-statistical theory}.
\newblock PhD thesis, Massachusetts Institute of Technology, 1993.

\bibitem{aharoni2000introduction}
A.~Aharoni, {\em Introduction to the {T}heory of {F}erromagnetism}, vol.~109.
\newblock Clarendon Press, 2000.

\bibitem{kikuchi1951theory}
R.~Kikuchi, ``A theory of cooperative phenomena,'' {\em Physical review},
  vol.~81, no.~6, pp.~988--1003, 1951.

\bibitem{weisstein2008stirling}
E.~W. Weisstein, ``Stirling's approximation,'' {\em MathWorld.
  http://mathworld.wolfram.com/StirlingsApproximation.html}, 2008.

\bibitem{kittel1966introduction}
C.~Kittel, {\em {I}ntroduction to {S}olid {S}tate {P}hysics}.
\newblock Wiley, New York, 6th~ed., 1986.

\bibitem{crangle1971magnetization}
J.~Crangle and G.~M. Goodman, ``The magnetization of pure iron and nickel,'' in
  {\em Proceedings of the Royal Society of London A: Mathematical, Physical and
  Engineering Sciences}, vol.~321, pp.~477--491, The Royal Society, 1971.

\bibitem{Note1}
Of course, as pointed out by Park \cite {park1968ensembles}, the use of a
  heterogeneous ensemble leads to the conclusion that knowledge of the state of
  the system, a bedrock of physical thought, is lost and all that can indeed be
  said is that the state of the ensemble and not that of the system is known.
  This is not the case for a homogeneous ensemble for which the state of the
  ensemble necessarily coincides with that of the system.

\bibitem{gyftopoulos2005thermodynamics}
E.~P. Gyftopoulos and G.~P. Beretta, {\em Thermodynamics: foundations and
  applications}.
\newblock Courier Corporation, 2005.

\bibitem{Note2}
The meaning of the term `complexions' corresponds to energy degeneracy used in
  the SEAQT framework.

\bibitem{casas1994nonequilibrium}
J.~Casas-V{\'a}zquez and D.~Jou, ``Nonequilibrium temperature versus
  local-equilibrium temperature,'' {\em Physical Review E}, vol.~49, no.~2,
  pp.~1040--1048, 1994.

\bibitem{pajda2001ab}
M.~Pajda, J.~Kudrnovsk{\`y}, I.~Turek, V.~Drchal, and P.~Bruno, ``Ab initio
  calculations of exchange interactions, spin-wave stiffness constants, and
  {C}urie temperatures of {F}e, {C}o, and {N}i,'' {\em Physical Review B},
  vol.~64, no.~17, p.~174402, 2001.

\bibitem{park1968ensembles}
J.~L. Park, ``The nature of quantum states,'' {\em American Journal of
  Physics}, vol.~36, p.~211, 1968.

\end{thebibliography}

\end{document}